\documentclass[iop]{emulateapj}
\usepackage{apjfonts}

\usepackage{epsfig,graphics,graphicx}
\pdfoutput=1

\newcommand{\arcm}{\hbox{$^\prime$}}

\newcommand{\degree}{\hbox{$^\circ$}}
\newcommand{\rosat}{\emph{ROSAT}}
\newcommand{\chandra}{\emph{Chandra}}
\newcommand{\xmm}{\emph{XMM-Newton}}
\newcommand{\xmms}{\emph{XMM}}
\newcommand{\asca}{\emph{ASCA}}

\newcommand{\arcs}{\mbox{\arcm\arcm}}
\newcommand{\Zsol}{\ensuremath{\mathrm{~Z_{\odot}}}}
\newcommand{\Lsol}{\ensuremath{\mathrm{~L_{\odot}}}}
\newcommand{\Msol}{\ensuremath{\mathrm{~M_{\odot}}}}
\newcommand{\Msolpyr}{\ensuremath{\mathrm{~M_{\odot}~yr^{-1}}}}

\newcommand{\NH}{\ensuremath{N_{\mathrm{H}}}} 

\newcommand{\s}{\ensuremath{\mbox{~s}}}
\newcommand{\ps}{\ensuremath{\s^{-1}}}
\newcommand{\cm}{\ensuremath{\mbox{~cm}}}
\newcommand{\pcmsq}{\ensuremath{\cm^{-2}}}
\newcommand{\pcmcu}{\ensuremath{\cm^{-3}}}
\newcommand{\km}{\ensuremath{\mbox{~km}}}
\newcommand{\Mpc}{\ensuremath{\mbox{~Mpc}}}
\newcommand{\pMpc}{\ensuremath{\Mpc^{-1}}}
\newcommand{\kmpspMpc}{\ensuremath{\km \ps \pMpc\,}}
\newcommand{\erg}{\ensuremath{\mbox{~erg}}}
\newcommand{\ergps}{\ensuremath{\erg \ps}}
\newcommand{\ergpspcmsq}{\ensuremath{\erg \ps \pcmsq}}

\newcommand{\kmps}{\ensuremath{\km \ps}}

\newcommand{\Hi}{\ensuremath{\mathrm{H}\textsc{i}}}

\newcommand{\Dtf}{\ensuremath{D_{25}}}

\begin{document}

\submitted{}
\received{2014 June 16}
\accepted{2014 July 26}

\title{Deep Chandra Observations of HCG~16 --- I. Active Nuclei, Star formation and Galactic Winds}
\author{E. O'Sullivan\altaffilmark{1}, A. Zezas\altaffilmark{1,2}, J.~M. Vrtilek\altaffilmark{1},  S. Giacintucci\altaffilmark{3,4}, M. Trevisan\altaffilmark{5}, L.~P. David\altaffilmark{1}, T.~J. Ponman\altaffilmark{6},\\ G.~A. Mamon\altaffilmark{7} and S. Raychaudhury\altaffilmark{8,6}}
\altaffiltext{1}{Harvard-Smithsonian Center for Astrophysics, 60 Garden
  Street, Cambridge, MA 02138, USA; eosullivan@cfa.harvard.edu}
\altaffiltext{2}{Physics Department and Institute of Theoretical \& Computational Physics, University of Crete, GR-71003 Heraklion, Crete, Greece}
\altaffiltext{3}{Department of Astronomy, University of Maryland, College
  Park, MD 20742-2421, USA}
\altaffiltext{4}{Joint Space Science Institute, University of Maryland, College Park, MD 20742-2421, USA}
\altaffiltext{5}{Instituto Nacional de Pesquisas Espaciais, Av. dos Astronautas 1758, 12227-010, S\~{a}o Jos\'{e} dos Campos, Brazil}
\altaffiltext{6}{School of Physics and Astronomy, University of Birmingham, Birmingham, B15 2TT, UK}
\altaffiltext{7}{Institut d'Astrophysique de Paris (UMR 7095 CNRS \& UMPC), 98 bis Bd Arago, F-75014 Paris, France}
\altaffiltext{8}{Department of Physics, Presidency University, 86/1 College Street, 700073 Kolkata, India}
\shorttitle{Deep \textit{Chandra} observations of HCG~16}
\shortauthors{O'Sullivan et al}

\begin{abstract}
  We present new, deep \chandra\ X-ray and \textit{Giant Metrewave Radio
    Telescope} 610~MHz observations of the spiral-galaxy-rich compact group
  HCG~16, which we use to examine nuclear activity, star formation and the
  high luminosity X-ray binary populations in the major galaxies. We
  confirm the presence of obscured active nuclei in NGC~833 and NGC~835,
  and identify a previously unrecognized nuclear source in NGC~838. All
  three nuclei are variable on timescales of months to years, and for
  NGC~833 and NGC~835 this is most likely caused by changes in accretion
  rate. The deep \chandra\ observations allow us to detect for the first
  time an Fe-K$\alpha$ emission line in the spectrum of the Seyfert 2
  nucleus of NGC~835. We find that NGC~838 and NGC~839 are both
  starburst-dominated systems, with only weak nuclear activity, in
  agreement with previous optical studies. We estimate the star formation
  rates in the two galaxies from their X-ray and radio emission, and
  compare these results with estimates from the infra-red and ultra-violet
  bands to confirm that star formation in both galaxies is probably
  declining after galaxy-wide starbursts were triggered $\sim$400-500~Myr
  ago. We examine the physical properties of their galactic superwinds, and
  find that both have temperatures of $\sim$0.8~keV. We also examine the
  X-ray and radio properties of NGC~848, the fifth largest galaxy in the
  group, and show that it is dominated by emission from its starburst.

\end{abstract}

\keywords{galaxies: groups: individual (HCG~16) --- galaxies: individual (NGC~833, NGC~835, NGC~838, NGC~839, NGC~848) --- galaxies: active --- galaxies: starburst --- X-rays: galaxies}

\section{Introduction}

Compact groups of galaxies provide an excellent natural laboratory for the
study of galaxy interactions and evolution. The group environment is
conducive to such interactions, with low velocity dispersions
($\lesssim$500\kmps), small galaxy separations, and galaxy densities
comparable to those seen in galaxy clusters. The repeated tidal
interactions between member galaxies in compact groups are likely to play a
role in a variety of processes involved in galaxy evolution, including gas
stripping, the triggering or quenching of star formation and galactic
winds, and the feeding and growth of the central supermassive black holes.
As most galaxies in the local universe reside in groups \citep{Ekeetal04}
and it is thought that many galaxies in clusters have been processed
through group environments in the past
\citep{Cappellarietal11,Mahajanetal12}, it is clearly important to
understand the role of the group environment in driving galaxy evolution .

The gas content and galaxy population of groups appears to be linked, with spiral-rich groups typically containing large reservoirs of \Hi\ and other cold gas, while elliptical dominated groups often have extended halos of hot, X-ray emitting gas, with neutral hydrogen restricted to spiral galaxies in the group outskirts \citep{Kilbornetal06}. Examination of X-ray faint, spiral-rich compact groups has led to the suggestion of an evolutionary sequence, with galaxy interactions stripping the \Hi\ from spiral galaxies to form intergalactic clouds and filaments or even a diffuse cold IGM \citep{VerdesMontenegroetal01,Johnsonetal07,Konstantopoulosetal10}. The redistribution of the \Hi\ component is accompanied by the transformation of some member galaxies from late to early-type, and in some cases by star formation. HCG~16, also known as Arp~318, is one of the best studied examples of a compact group in the late stages of this evolutionary sequence.

\begin{figure*}
\centerline{
\includegraphics[width=1.05\columnwidth,bb=36 175 576 616]{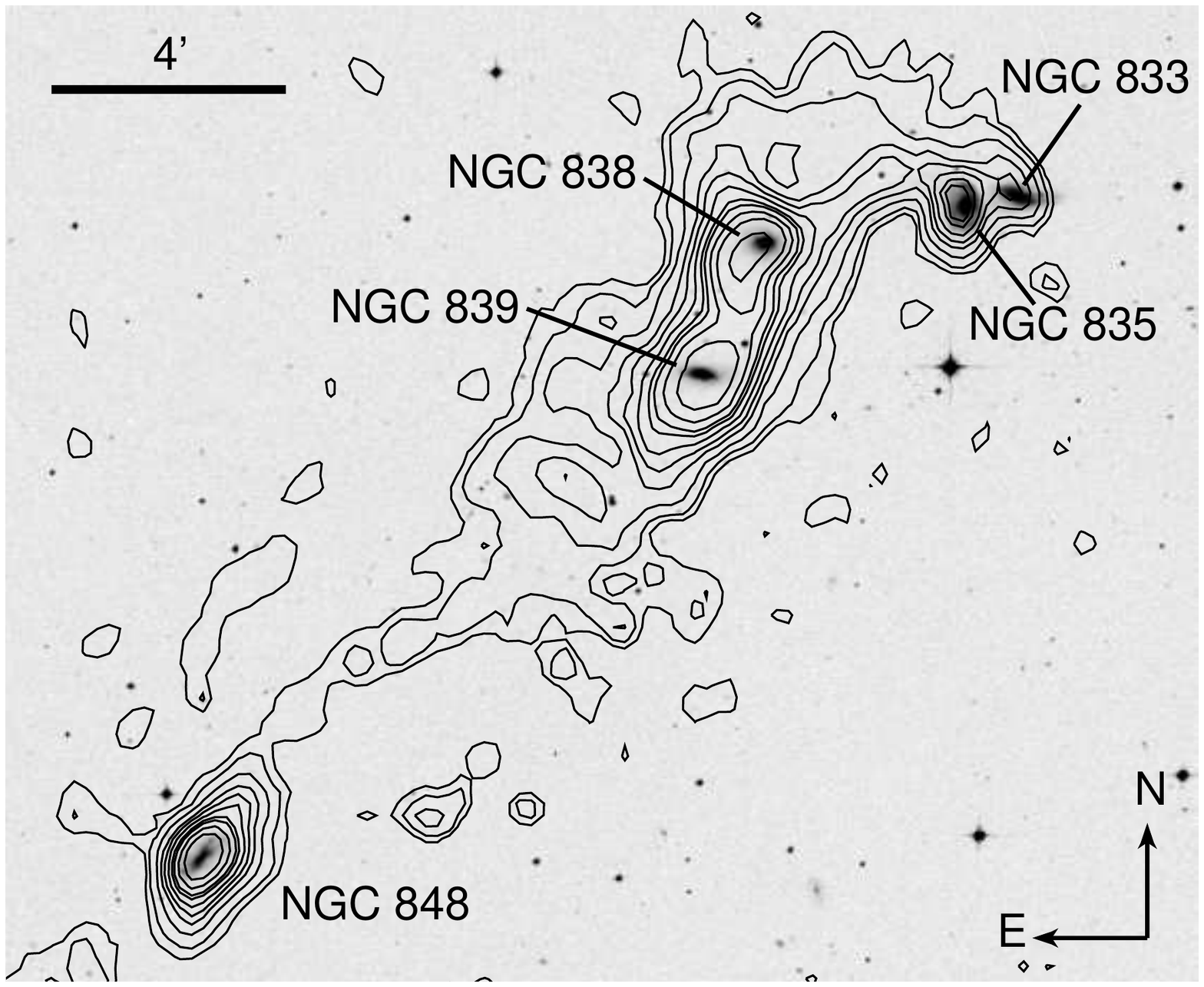}
\includegraphics[width=1.05\columnwidth,bb=36 175 576 616]{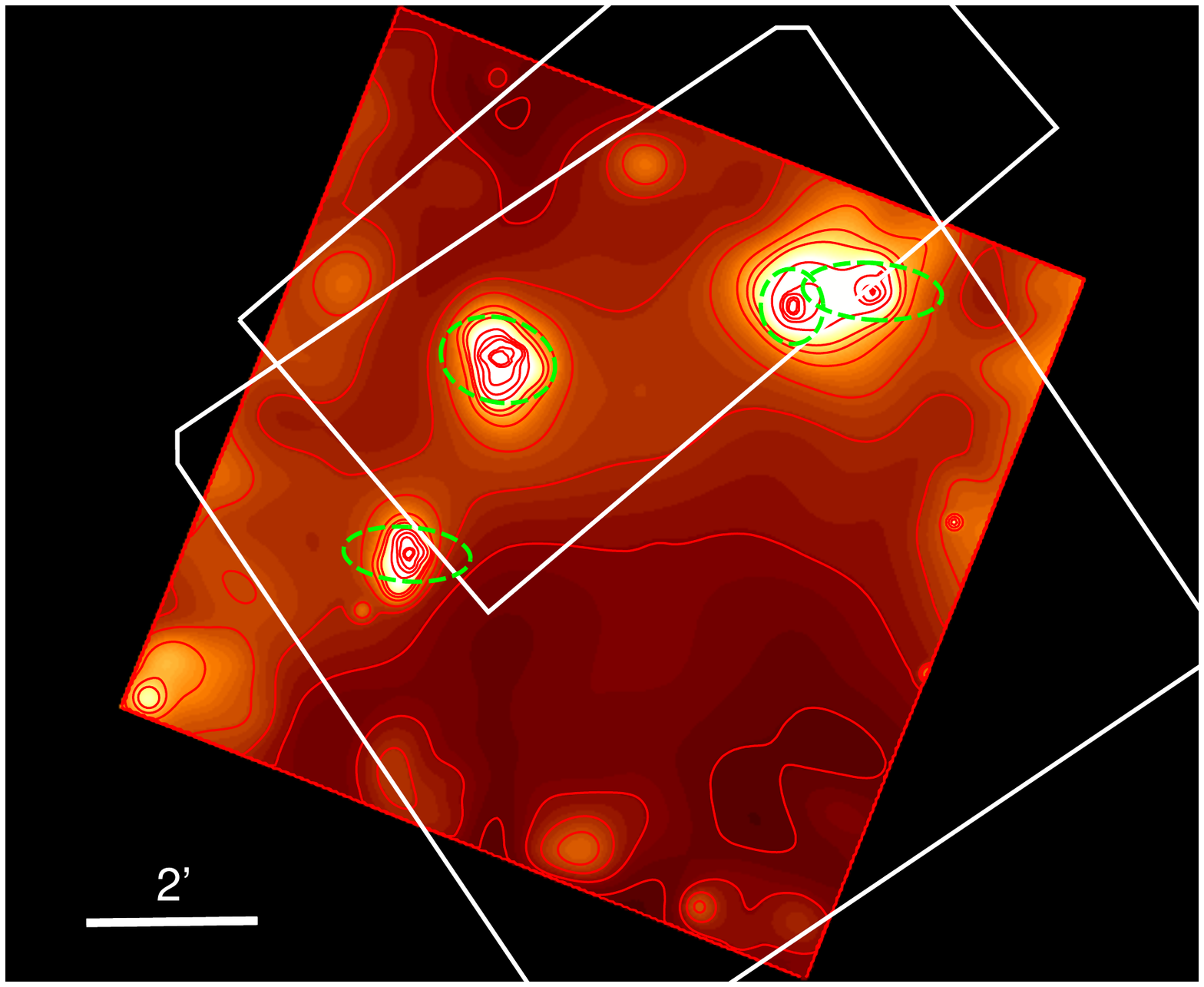}
}
\caption{\label{fig:Xopt} \textit{Left:} Digitized Sky Survey 2 (DSS2) $R$-band image of the five largest galaxies in HCG~16, with the four galaxies originally identified as a compact group to the northwest. VLA \Hi\ contours from Verdes-Montenegro et al. (2014, in prep.) are overlaid, with levels N(\Hi)$\simeq$10,20,40,65,85,110,140,160,200,250,350,450,570$\times$10$^{-19}$\pcmsq. \textit{Right:} Adaptively smoothed \chandra\ 0.5-2~keV image using data from the S3 CCD in all five observations. Contours are overlaid in red to help elucidate the distribution of diffuse emission. Dashed ellipses indicate the \Dtf\ contours of the four main galaxies. Cyan regions indicate the active sections of the S3 CCD in ObsID 923 (roughly square) and 10394 (rectangular). The two images have the same orientation, but different scales, as indicated by the scalebars.}
\end{figure*}

HCG~16 was originally identified \citep{Hickson82} as a compact group of
four spiral galaxies, NGC~833 (HCG~16B), NGC~835 (A), NGC~838 (C) and
NGC~839 (D). Figure~\ref{fig:Xopt} includes a Digitized Sky Survey image
showing the relative positions of the group members. Later observations
increased the number of members to seven, including one large spiral
galaxy, NGC~848, of comparable luminosity to the original four
\citep{Ribeiroetal98}. A search in the NASA/IPAC Extragalactic Database
(NED\footnote{The NASA/IPAC Extragalactic Database (NED) is operated by the
  Jet Propulsion Laboratory, California Institute of Technology, under
  contract with the National Aeronautics and Space Administration.}) finds
three additional galaxies within 30\arcm\ and in the velocity range
3800-4100\kmps, suggesting a more dispersed halo of dwarf galaxies
surrounding the core. All five major galaxies host AGN and/or starbursts
\citep[e.g.,][]{Martinezetal10,DeCarvalhoCoziol99}, and an ongoing or
recent interaction between NGC~833 and NGC~835 has warped their galaxy
disks and produced a dusty tidal arm extending east from NGC~835 toward
NGC~838 \citep{Konstantopoulosetal13}. Both \citet{Mulchaeyetal03} and
\citet{OsmondPonman04} find spiral fractions for the group $<$1 ($f_{\rm
  sp}$=0.86, and $f_{\rm sp}$=0.83 respectively), but in both cases this
primarily reflects the classification of NGC~839 as an S0, which seems
misleading given its current high rate of star formation. We consider all
five major galaxies in the group to be late-type.

Neutral hydrogen mapping of the group (see Figure~\ref{fig:Xopt}) revealed a
$\sim$20\arcm\ long complex filament of cold gas surrounding the four
original members of the group and linking them to NGC~848
\citep{VerdesMontenegroetal01}, almost certainly as the result of tidal
interactions between group members. The total mass of \Hi\ in the group is
$>$2.63$\times$10$^{10}$\Msol, and Verdes-Montenegro et al. estimate that
the group is $<$30\% \Hi-deficient. The four original member galaxies are
$\sim$50-80\% deficient, while NGC~848 is only $\sim$7\% deficient. This
suggests that the majority of the intergalactic \Hi\ originated in the four
main galaxies, perhaps being transported out into the IGM by interactions
among them, and then drawn into its current morphology by a close passage
of NGC~848. \citet{Borthakuretal10} show that the \Hi\ velocity
distribution covers the range $\sim$3650-4100\kmps, confirming its
association with the major member galaxies.

HCG~16 was first detected in X-rays  using the \textit{Einstein} observatory \citep{Bahcalletal84}. More detailed studies with \rosat\ were able to separate emission from the galaxies and diffuse inter-galactic gas \citep{Ponmanetal96,DosSantosMamon99} with the brightest emission around and between the four main galaxies. First light data from \xmm\ were used to examine the four main galaxies, providing evidence of a combination of star formation and AGN emission from NGC~833, NGC~835 and NGC~839, and classifying NGC~838 as a pure starburst \citep[][hereafter TRP01]{Turneretal01special}. A short (12.5~ks) \chandra\ observation in cycle~1 was unable to improve on the characterization of the galaxy emission \citep{GonzalezMartinetal06} but did provide evidence of a bridge of diffuse emission linking NGC~833 and NGC~835 \citep[][see also the unpublished image previously produced from this data by Mamon \& Lima Neto\footnote{
http://dx.doi.org/10.6084/m9.figshare.971366}]{Jeltemaetal08}. 

Optical spectroscopic studies of the starburst galaxies NGC~838 and NGC~839 have provided detailed characterization of these two galaxies and their outflowing galactic winds. In NGC~839 the wind has formed a biconical polar outflow with velocity $\sim$250\kmps, while the galaxy is dominated by a rapidly rotating $\sim$400~Myr old stellar component \citep{Richetal10}. The winds appear to be shock excited, and Rich et al. argue that emission probably arises from these shocks rather than from an AGN. NGC~838 appears to have undergone a galaxy-wide starburst episode $\sim$500~Myr ago, although rapid star formation is now found only in the galaxy core \citep{Vogtetal13}. The asymmetric photoionized winds have inflated bubbles north and south of the galaxy, above and below the galactic disk, with sizes of $\sim$5~kpc and 7~kpc ($\sim$18\arcs\ and 25\arcs) for the north and south bubbles respectively, and likely ages of 5-50~Myr. Highly redshifted gas emission in the outer part of the larger southern bubble indicates that it is probably leaking material.  

In this paper we use new, deep \chandra\ observations to examine the major
galaxies of HCG~16, with the goal of studying their point source
populations, star formation and nuclear activity. A study of the diffuse
emission in the group, and its relation to the major galaxies is presented
in \citet[hereafter paper II]{OSullivanetal14c_special}. We adopt a Galactic
hydrogen column density of \NH=2.56$\times$10$^{20}$\pcmsq\ for the four
original group member galaxies and the surrounding diffuse emission
\citep[taken from the Leiden/Argentine/Bonn survey,][]{Kalberlaetal05}. For
NGC~848 we adopt a hydrogen column of \NH=2.75$\times$10$^{20}$\pcmsq. All
fluxes and luminosities are corrected for Galactic absorption. We adopt a
redshift of $z$=0.0132 for the group \citep{Hicksonetal92}. A
redshift-independent distance measurement is available for one of the five
major galaxies, a Tully-Fisher distance of 56.5~Mpc for NGC~848
\citep{Theureauetal07}. This is consistent, within errors, with
redshift-based estimates for all five of the galaxies, correcting for
infall toward the Virgo cluster, great attractor and Shapley Supercluster,
for a cosmology with $H_0$=70\kmpspMpc.  We therefore adopt this distance
estimate for the group as a whole, which gives an angular scale of
1\arcs=273~pc.

\section{Observations and Data Analysis}

\subsection{Chandra}
\label{sec:obs}

\begin{deluxetable*}{lccccccc}
\tablewidth{0pt}
\tablecaption{\label{tab:obs}Summary of \chandra\ observations of HCG~16}
\tablehead{
\colhead{ObsID} & \colhead{P.I.} & \colhead{Observation date} & \colhead{mode} & \colhead{Frame time} & \colhead{subarray} & \colhead{Roll angle} & \colhead{Cleaned exposure} \\
\colhead{} & \colhead{} & \colhead{} & \colhead{} & \colhead{(s)} & \colhead{} & \colhead{(\degree)} & \colhead{(s)}
}
\startdata
\dataset[ADS/Sa.CXO#obs/923]{923}     & Mamon     & 2000 Nov 16 & F  & 3.2 & N   & 325.98 & 12565 \\ 
\dataset[ADS/Sa.CXO#obs/10394]{10394} & Alexander & 2008 Nov 23 & VF & 1.5 & 1/2 & 319.47 & 13824 \\
\dataset[ADS/Sa.CXO#obs/15181]{15181} & Vrtilek   & 2013 Jul 16 & VF & 3.1 & N   & 111.35 & 49457 \\
\dataset[ADS/Sa.CXO#obs/15666]{15666} & Vrtilek   & 2013 Jul 18 & VF & 3.1 & N   & 111.35 & 29714 \\
\dataset[ADS/Sa.CXO#obs/15667]{15667} & Vrtilek   & 2013 Jul 21 & VF & 3.1 & N   & 111.35 & 58335
\enddata
\end{deluxetable*}

HCG~16 has been observed five times by the \chandra\ ACIS instrument,
briefly in cycles~1 and 10, and most recently in cycle~14 for a total of
137.5~ks. Table~\ref{tab:obs} summarises the observational setup of the
five exposures. A summary of the \chandra\ mission and instrumentation
can be found in \citet{Weisskopfetal02}. In all five observations the S3
CCD was placed at the focus of the telescope. ObsID~10394 was performed
using a 1/2 subarray, with a short (1.5~s) frame time, and with only the
ACIS-S3 CCD switched on. The three observations in cycle~14 were performed
with the same roll angle. The four original group members are located on
the S3 CCD in four of the five observations, but ObsID~10394 covers only NGC~835 and NGC~838. NGC~848 is only visible on the S1 CCD of ObsID~923.

We have reduced the data from all five pointings using CIAO 4.6.1
\citep{Fruscioneetal06} and CALDB 4.5.9 following techniques similar to
those described in \citet{OSullivanetal07} and the \chandra\ analysis
threads\footnote{http://asc.harvard.edu/ciao/threads/index.html}.  The
level 1 event files were reprocessed, bad pixels and events with \asca\
grades 1, 5 and 7 were removed, and the cosmic ray afterglow correction was
applied. Very Faint mode cleaning was applied to all observations except
ObsID~923. The data were corrected to the appropriate gain map, the
standard time-dependent gain and charge-transfer inefficiency (CTI)
corrections were made, and background light curves were produced. None of
the observations suffered from significant background flaring. A comparison
of 0.5-0.7~keV and 2.5-5~keV light curves shows no significant difference
between the two bands, indicating that the observations are not affected by
solar wind charge exchange emission. In general all five observations were
combined for imaging analysis, but spectra were extracted separately and
fitted simultaneously.

Point source identification was performed using the \textsc{ciao} task
\textsc{wavdetect}, with a detection threshold of 10$^{-6}$, chosen to
ensure that the task detects $\lesssim$1 false source in the S3 field of
view, working from a 0.3-7.0 keV image and exposure map from the five
observations combined. We defined source apertures based on the position of
each source as reported by \textsc{wavdetect}. In the case of confused or
extended sources we adjusted these centroids based on the peak of the
emission in the 2-7~keV band. Ellipticities and position angles from
\textsc{wavdetect} were retained, but the axes of each source region were
reduced to approximately the 90\% encircled energy radius, with a minimum
radius of 1\arcs. For sources outside the bodies of the galaxies we used
the regions reported by \textsc{wavdetect}, which satisfied the $>$90\%
encircled energy limit. The resulting regions were used to examine source
properties, and to exclude them from spectral fits to the diffuse emission.

Spectra were extracted from each dataset using the \textsc{specextract}
task. When examining diffuse emission, background spectra were drawn from
blank-sky event lists, scaled to match the data in the 9.5-12.0 keV band.
For point sources, local background spectra were used, typically from an
elliptical annulus with radius 1.5-3 times that of the source extraction
region. Where the source was partially surrounded by diffuse emission (as
in NGC~838), background regions were chosen to include a representative
fraction of that diffuse emission.  Spectral fitting was performed in XSPEC
12.8.1.  Abundances were measured relative to the abundance ratios of
\citet{GrevesseSauval98}. 1$\sigma$ uncertainties for one interesting
parameter are reported for all fitted values.

\subsection{Very Large Array and Giant Metrewave Radio Telescope}
\begin{deluxetable*}{lcccccccc}[!t]
\tablewidth{0pt}
\tablecaption{\label{tab:Robs}Summary of radio observations of HCG~16}
\tablehead{
\colhead{Observatory} & \colhead{project} & \colhead{Observation date}& \colhead{Frequency} & \colhead{Bandwidth} & \colhead{Integration time} & \colhead{HPBW} & \colhead{Pos. Angle} & \colhead{rms} \\
\colhead{} & \colhead{} & \colhead{} & \colhead{(MHz)} & \colhead{(MHz)} & \colhead{(hr)} & \colhead{(\arcs$\times$\arcs)} & \colhead{(\degree)} & \colhead{($\mu$Jy beam$^{-1}$)}
}
\startdata
VLA  & AW500  & 1999 Jan 13-14 & 1402 & 6.3 & 5 & 25.0$\times$18.1 & -6.64 & 200 \\
GMRT & 17\_026 & 2009 Nov 21    & 610  & 32  & 6 & 5.6$\times$5.4   & -3.0  & 60              
\enddata
\end{deluxetable*}

We analysed 1.4~GHz data retrieved from the \textit{Very Large Array} (VLA) public
archive. The observations were performed in C array, in spectral line mode
with 64 channels centred at 1402~MHz for a total on-source exposure of
$\sim$5~hrs. Further details of the observation are shown in
Table~\ref{tab:Robs}. We used the NRAO Astronomical Image Processing System
(\textsc{aips}) package for the data reduction and analysis. Data
calibration and imaging were carried out following the standard procedure
(Fourier transform, clean and restore). Phase-only self-calibration was
applied to remove residual phase variations and improve the quality of the
image.  The final image has an angular resolution of 25$\times$18.1\arcs\
and an rms noise level (1$\sigma$) of 0.2~mJy beam$^{-1}$.

We also observed the group with the \textit{Giant Metrewave Radio Telescope} (GMRT) in dual-frequency 610/235~MHz mode (project 17$\_$026, P.I. E. O'Sullivan). The observations were again analysed using \textsc{aips}, with phase self-calibration, following the methods described in \citet{Giacintuccietal11}. The 235~MHz data suffer from calibration problems, and we therefore excluded them from further analysis. The 610~MHz data are essentially free of radio frequency interference (RFI), but are dynamic range limited owing to a powerful nearby source whose side lobes affect the region of the group. This problem becomes more severe in lower resolutions, and we therefore use only the full resolution (5.6$\times$5.4\arcs) image which has an rms noise level (1$\sigma$) of 0.06~mJy~beam$^{-1}$. Further details of the 610~MHz observation are shown in Table~\ref{tab:Robs}.

\section{Point Sources}
\label{sec:PS}
Given its distance, we only expect to be able to detect relatively bright
point sources in HCG~16. We use the \chandra\ Portable Interacting
Multi-Mission Simulator \citep[PIMMS;][]{Mukai93} to estimate the detection
limit of the combined 2013 observations, requiring a minimum of 10 counts
(0.5-7~keV) for detection and adopting a powerlaw model with $\Gamma$=1.7
and Galactic absorption, and find a limit of
$L_{0.5-7}\geq$2.1$\times$10$^{38}$\ergps\
($F_{0.5-7}\geq$5.5$\times$10$^{-16}$\ergpspcmsq). This is only about a
factor of five below our adopted threshold for ultra--luminous X--ray point
sources (ULXs), $L_{0.5-8}\geq$10$^{39}$\ergps\
\citep[e.g.,][]{Swartzetal04}.  Since these are star forming galaxies, we
expect most sources in the galaxies (excluding AGN) to be high mass X-ray
binaries (HMXBs), formed from massive, short-lived stars.

Of the 30 point sources identified in the ACIS-S3 field of view, 18 fall
within the \Dtf\ ellipses of the four major galaxies, or in the tidal
structures around NGC~833 and NGC~835. Table~\ref{tab:sources} summarises
the locations of these 18 sources, the regions used to measure their
fluxes, and the number of net counts in each source in a co-added image of
all five observations. As expected, many of the sources in the starburst
galaxy NGC~838 are clustered in the galaxy core, where star formation is
ongoing.  The other three galaxies contain only 1-4 sources each, and there
is no clear correlation between source position and galaxy structure. We
initially extracted spectra for each source, and estimated their background
subtracted count rates using the \textsc{dmextract} task. The three sources
with the highest fluxes ($>$700 net 0.5-7~keV counts) are associated with
the nuclei of NGC~833, NGC~835 and NGC~839 (HCG~16B, A, and D). Detailed
fits to these sources and their surrounding diffuse emission will be
described in \textsection~\ref{sec:AGNdiff}. A further four sources had a
sufficient number of counts in individual exposures to allow fitting of
simple absorbed powerlaw and absorbed APEC thermal models, to help
determine their origin. These fits are described later in
\textsection~\ref{sec:PSfits}.

The remaining eleven sources were too faint for individual spectral
fitting. In order to estimate luminosities from their count rates we
required a conversion factor, including corrections for Galactic absorption
and for the fraction of flux scattered outside the extraction region by the
\chandra\ point spread function (PSF). As the responses calculated for each
source during spectral extraction include a correction for PSF scattering,
this conversion factor can be found by folding a standard source spectral
model through each set of responses and determining the expected flux for a
fixed count rate. Extragalactic X-ray binaries are commonly modelled using
spectrally hard models such as a $\Gamma$=1.7 powerlaw
\citep[e.g.,][]{Smithetal12}. To test the suitability of such a model we
co-added the spectra and responses of the eleven faint sources using the
\textsc{combine\_spectra} task to create three stacked spectra, one each
for ObsID 923, 10394, and the 2013 observations (15181, 15666 and 15667).
These spectra were then simultaneously fitted using an absorbed powerlaw
model with hydrogen column fixed at the Galactic value. The best fit has
$\Gamma$=1.69$\pm$0.14, in agreement with expectations, and reduced
$\chi^2$=1.245 for 22 degrees of freedom. We therefore adopt the
$\Gamma$=1.7 powerlaw to calculate the counts-to-flux conversion factors
for each source in each observation. The resulting fluxes are listed in
Table~\ref{tab:variable}. Where sources are detected at less than 3$\sigma$
significance, we calculate an upper limit based on the integer number of
counts required to produce a 3$\sigma$ detection, taking into account the
scaled local background.

\begin{figure*}
\includegraphics[width=\textwidth,bb=30 550 570 760]{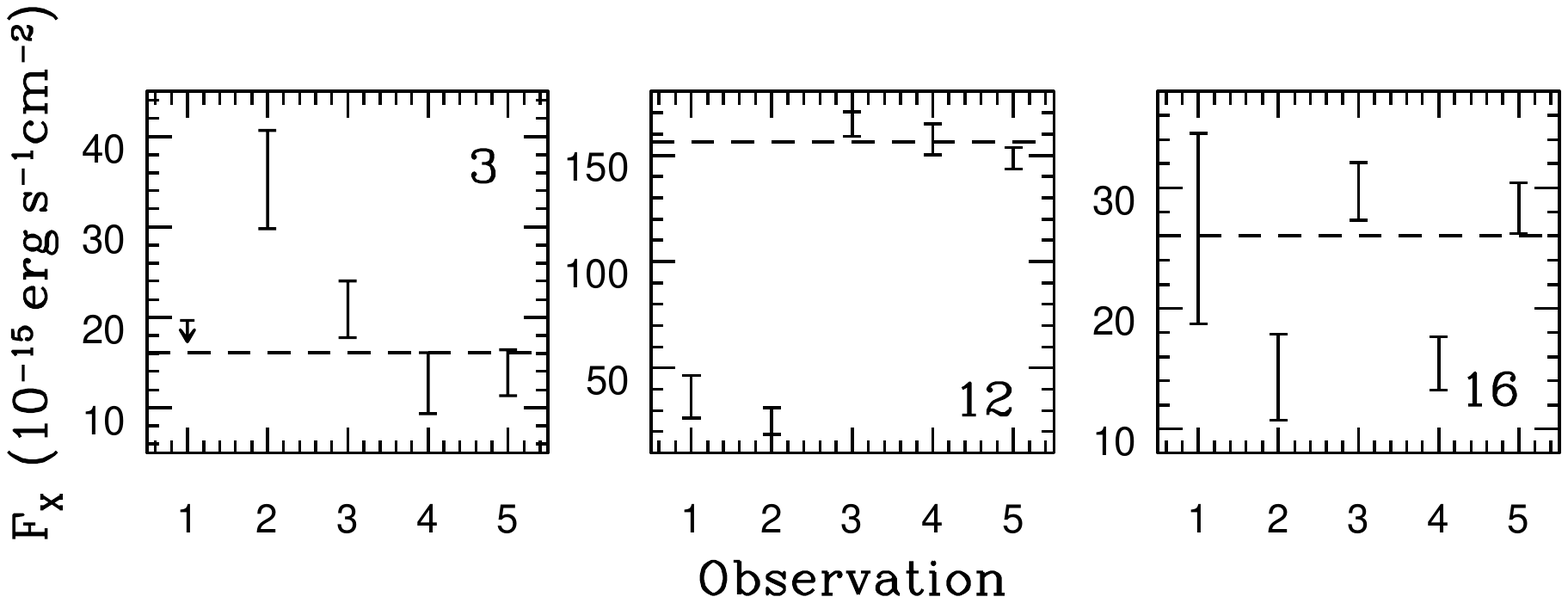}
\caption{\label{fig:var}Variability plots for those sources whose 0.5-7~keV fluxes appear to change over time, with observations in chronological order from ObsID 923 (bin 1) to 15667 (bin 5). 1$\sigma$ error bounds on the counts-based fluxes from the five observations, or 3$\sigma$ upper limits are shown for each source, with the mean flux in the 2013 observations marked by a dashed line. Each plot is labelled with the source number.}
\end{figure*}

We also estimate the mean flux in the three 2013 observations combined;
individual sources can be detected in a stacked image yet be undetected in
each individual observation, and combining the three most recent
observations allows us to estimate a flux without trying to combine counts
from the earlier observations which have very different conversion factors.
Three of the sources show evidence of variability between observations at
$>$3$\sigma$ significance. Figure~\ref{fig:var} shows the
0.5-7~keV fluxes for these sources in each observation. We note
that when classifying sources as potential ULXs, we estimate a factor of
1.075 increase in flux between the 0.5-7~keV and 0.5-8~keV bands, so
sources with $L_{0.5-7}$$\geq$9.32$\times$10$^{38}$\ergps\ are bright
enough to be ULXs if they are located within HCG~16.

\begin{deluxetable*}{lcccccc}
\tablewidth{0pt}
\tablecaption{\label{tab:sources}List of point sources and their properties}
\tablehead{
\colhead{Source} & \colhead{R.A.} & \colhead{Dec.} & \colhead{Radii$^a$} & \colhead{p.a.$^a$} & \colhead{net counts$^b$} & \colhead{Notes} \\
\colhead{} & \colhead{(J2000)} & \colhead{(J2000)} & \colhead{(\arcs)} & \colhead{(\degree)} & \colhead{(cnt.~s$^{-1}$)} & \colhead{}
}
\startdata
\multicolumn{7}{l}{\textit{NGC~839}} \\
1 & 2$^{\text{h}}$09${\text{m}}$42.754$^{\text{s}}$ & -10$^\circ$11\arcm 02.40\arcs & 2.09,1.65 & 24.7 & 735.3$\pm$29.4 & nucleus\\
\hline
\multicolumn{7}{l}{\textit{NGC~838}} \\
2 & 2$^{\text{h}}$09${\text{m}}$37.062$^{\text{s}}$ & -10$^\circ$08\arcm 55.91\arcs & 2.19,1.00 & 50.3 & 27.8$\pm$5.5 & \\
3 & 2$^{\text{h}}$09${\text{m}}$38.531$^{\text{s}}$ & -10$^\circ$08\arcm 48.26\arcs & 1.45,1.30 & 139.9 & 274.0$\pm$24.3 & ULX or nuclear source\\
4 & 2$^{\text{h}}$09${\text{m}}$37.984$^{\text{s}}$ & -10$^\circ$08\arcm 48.83\arcs & 1.61,1.00 & 49.2 & 93.3$\pm$11.0 & ULX\\
5 & 2$^{\text{h}}$09${\text{m}}$38.128$^{\text{s}}$ & -10$^\circ$08\arcm 45.71\arcs & 1.19,1.14 & 155.5 & 24.4$\pm$12.5 & ULX, 1.44\arcs\ from CXO J020938.1-100847\\
6 & 2$^{\text{h}}$09${\text{m}}$38.449$^{\text{s}}$ & -10$^\circ$08\arcm 44.87\arcs & 1.27,1.20 & 151.3 & 101.4$\pm$19.0 & probably thermal, 1.75\arcs\ from SN 2005H \\
7 & 2$^{\text{h}}$09${\text{m}}$39.237$^{\text{s}}$ & -10$^\circ$08\arcm 44.41\arcs & 1.57,1.31 & 15.4 & 36.2$\pm$7.4 & \\
8  & 2$^{\text{h}}$09${\text{m}}$38.340$^{\text{s}}$ & -10$^\circ$08\arcm 34.93\arcs & 1.32,1.10 & 25.0 & 21.0$\pm$5.7 & \\
9  & 2$^{\text{h}}$09${\text{m}}$40.591$^{\text{s}}$ & -10$^\circ$08\arcm 26.72\arcs & 1.12,1.00 & 24.8 & 15.5$\pm$4.4 & \\
11 & 2$^{\text{h}}$09${\text{m}}$38.118$^{\text{s}}$ & -10$^\circ$08\arcm 19.64\arcs & 1.26,1.05 & 140.4 & 34.3$\pm$6.1 & possible background AGN in SDSS~J020938.10-100819.4\\
17 & 2$^{\text{h}}$09${\text{m}}$37.962$^{\text{s}}$ & -10$^\circ$08\arcm 40.53\arcs & 1.72,1.20 & 147.4 & 58.8$\pm$9.6 & possibly thermal, CXO J020937.9-100840 \\
\hline
\multicolumn{7}{l}{\textit{NGC~833 / NGC~835 complex}} \\
10 & 2$^{\text{h}}$09${\text{m}}$22.900$^{\text{s}}$ & -10$^\circ$08\arcm 24.93\arcs & 1.45,1.45 & 174.6 & 116.6$\pm$11.1 & ULX in tidal zone, CXO J020922.8-100824 \\ 
12 & 2$^{\text{h}}$09${\text{m}}$24.613$^{\text{s}}$ & -10$^\circ$08\arcm 09.50\arcs & 1.61,1.70 & 82.3 & 2361.3$\pm$50.7 & NGC~835 nucleus\\
13 & 2$^{\text{h}}$09${\text{m}}$21.099$^{\text{s}}$ & -10$^\circ$08\arcm 03.15\arcs & 1.42,1.13 & 154.8 & 33.4$\pm$6.9 & ULX\\
14 & 2$^{\text{h}}$09${\text{m}}$20.861$^{\text{s}}$ & -10$^\circ$07\arcm 59.41\arcs & 1.53,1.34 & 86.7 & 718.4$\pm$27.7 & NGC~833 nucleus\\
15 & 2$^{\text{h}}$09${\text{m}}$20.427$^{\text{s}}$ & -10$^\circ$07\arcm 48.08\arcs & 1.64,1.20 & 37.0 & 17.9$\pm$4.5 & \\
16 & 2$^{\text{h}}$09${\text{m}}$27.578$^{\text{s}}$ & -10$^\circ$07\arcm 46.72\arcs & 1.39,1.51 & 166.1 & 421.1$\pm$20.8 & possible background AGN, CXO J020927.6-100746 \\
 & & & & & & in SDSS~J020927.57-100746.2 \\
18 & 2$^{\text{h}}$09${\text{m}}$20.773$^{\text{s}}$ & -10$^\circ$07\arcm 46.18\arcs & 2.44,1.00 & 178.1 & 11.5$\pm$3.9 & 
\enddata
\tablecomments{
$^a$ Radii and position angles of elliptical regions used to extract spectra and numbers of counts. $^b$ Counts in 0.5-7~keV band, summed over all observations with source in field of view. 
}
\end{deluxetable*}

\begin{deluxetable*}{lccccccc}
\tablewidth{0pt}
\tablecaption{\label{tab:variable}Point source fluxes in each observation and the stacked 2013 observations}
\tablehead{
\colhead{Source} & \multicolumn{6}{c}{L$^{cts}_{0.5-7}$ (10$^{38}$\ergps )} & \colhead{Variable?} \\
\colhead{} & \colhead{923} & \colhead{10394} & \colhead{15181} & \colhead{15666} & \colhead{15667} & \colhead{15XXX$^a$} & \colhead{} \\
}
\startdata
1  &    236.13$\pm$25.20 & -          &   316.83$\pm$23.22 &   286.23$\pm$29.56 &   303.88$\pm$18.86 &   311.05$\pm$13.40 & N \\ 
2  &     $<$33.12 & -          &    $<$8.10 &    $<$17.42 &    13.52$\pm$3.55 &    8.28$\pm$1.76 & - \\

3  &     $<$75.05 &   134.64$\pm$20.81 &   79.76$\pm$11.97 &   48.64$\pm$12.93 &   53.11$\pm$9.67 &   61.57$\pm$6.57 & $>$3$\sigma$\\

4  &     $<$44.12 &    $<$24.71 &    23.94$\pm$5.55 &    $<$20.89 &    34.78$\pm$5.85 &    27.14$\pm$3.47 & N \\ 

5  &     $<$58.25 &    $<$70.62 &    $<$33.76 &    $<$4.43 &    $<$29.87 &    10.91$\pm$6.18 & - \\

6  &     $<$73.49 &    $<$47.13 &    $<$23.03 &    $<$28.68 &    $<$20.85 &    22.48$\pm$4.92 & - \\

7  &     $<$48.51 &    $<$24.75 &    $<$10.12 &    $<$15.39 &    $<$9.01 &    5.41$\pm$1.65 & - \\

8  &     $<$54.16 &    $<$48.78 &    $<$9.78 &    $<$14.90 &    $<$8.25 &    5.69$\pm$1.54 & - \\

9  &     $<$78.22 &    $<$22.61 &    $<$10.47 &    $<$17.42 &    $<$9.74 &    4.70$\pm$1.47 & - \\

10 &     $<$69.97 &    -             &    27.11$\pm$4.53 &    28.15$\pm$5.91 &    30.48$\pm$4.43 &   28.77$\pm$2.80 & N \\

11 &     $<$48.01 &    $<$23.76 &    $<$7.98 &    $<$16.50 &    $<$6.76 &    7.71$\pm$1.60 & - \\

12 &   139.48$\pm$37.98 &   95.77$\pm$24.07 &  628.75$\pm$22.00 &  601.34$\pm$27.64 &  567.72$\pm$19.27 &  597.10$\pm$12.85 & $>$16$\sigma$\\

13 &    $<$139.53 &    -              &    $<$10.12 &    16.84$\pm$5.29 &    9.34$\pm$2.87 &    9.72$\pm$1.95 & N \\ 

14 &  302.29$\pm$69.55 &   -              &  158.36$\pm$11.11 &  203.09$\pm$16.13 &  196.44$\pm$11.36 &  184.16$\pm$7.16 & N \\ 

15 &    $<$99.99 &    -              &    $<$7.75 &    $<$12.91 &    $<$6.61 &    4.81$\pm$1.23 & - \\

16 &  101.76$\pm$30.23 &   54.63$\pm$13.57 &  113.51$\pm$9.09 &   58.91$\pm$8.40 &  108.12$\pm$8.07 &   99.29$\pm$5.06 & $>$3$\sigma$\\

17 &    $<$42.13 &    $<$29.22 &   17.56$\pm$4.50 &    $<$19.56 &   11.78$\pm$3.64 &   14.28$\pm$2.58 & N \\ 

18 &   $<$90.41 &    -             &    $<$8.67 &    $<$15.13 &    $<$7.10 &    0.54$\pm$0.26 & - 

\enddata
\tablecomments{Upper limits are shown for sources detected at $<$3$\sigma$ significance in a particular observation. $^a$: Column 7 shows the mean luminosity in the 2013 observations 15181, 15666 and 15667 co-added.}
\end{deluxetable*}

\subsection{Notes on individual point sources}
\label{sec:PSfits}
For sources 3, 6, 10 and 16 we are able to fit simple absorbed thermal plasma or powerlaw models. Spectra from datasets 923 and 10394 were not included in cases where they contain only a handful of counts for each source. The results of these fits are shown in Table~\ref{tab:bright}.

\begin{deluxetable*}{lccccccc}
\tablewidth{0pt}
\tablecaption{\label{tab:bright}Spectral fits to the four point sources with highest fluxes not already known to be AGN}
\tablehead{
\colhead{Source} & \multicolumn{4}{c}{Thermal model} & \multicolumn{3}{c}{Powerlaw model} \\
\colhead{} & \colhead{kT} & \colhead{Abund.} & \colhead{$L_{0.5-7}$} & \colhead{red. $\chi^2$/d.o.f$^a$} & \colhead{$\Gamma$} & \colhead{$L_{0.5-7}$} & \colhead{red. $\chi^2$/d.o.f$^a$} \\
\colhead{} & \colhead{(keV)} & \colhead{(\Zsol)} & \colhead{(10$^{38}$\ergps)} & \colhead{} & \colhead{} & \colhead{(10$^{38}$\ergps)} & \colhead{}
}
\startdata
3 & $>$30.98 & 0.3$^b$ & 90.37$^{+6.61}_{-11.29}$ & 1.859/17 & 0.86$\pm$0.23 & 105.88$^{+14.63}_{-13.71}$ & 1.688/17 \\[+0.5mm]
6 & 1.44$^{+0.89}_{-0.20}$ & 0.3$^b$ & 15.39$^{+5.92}_{-4.32}$ & 1.63/4 & 2.85$^{+1.41}_{-1.54}$ & 20.66$^{+8.59}_{-5.81}$ & 2.05/4 \\[+0.5mm]
10 & 3.48$^{+2.73}_{-1.16}$ & 0.3$^b$ & 31.93$^{+4.24}_{4.16}$ & 0.532/6 & 2.00$\pm$0.30 & 34.60$^{+4.66}_{-4.58}$ & 0.60/6 \\[+0.5mm]
16 & 4.27$^{+1.30}_{-0.84}$ & 0.3$^b$ & 97.55$^{+7.83}_{-7.49}$ & 0.595/13 & 1.80$^{+0.12}_{-0.11}$ & 104.31$^{+7.03}_{-6.99}$ & 0.581/13 
\enddata
\tablecomments{$^a$The reduced $\chi^2$ of the fit over its degrees of freedom. $^b$ Parameter fixed during fitting.}
\end{deluxetable*}

Sources 3 and 6 are located close together in the core of NGC~838. Source 6 is best fitted by a thermal plasma model with temperature $\sim$1.4~keV, somewhat hotter than the surrounding diffuse thermal emission, but not unphysically so. It has a relatively steep spectral index when fitted with a powerlaw. Given that NGC~838 hosts an extended starburst wind with a good deal of clumpiness in its X--ray structure, it seems likely that this source is in fact a cloud of hot gas. 

By contrast, in source 3 the thermal model has an unconstrained high temperature while the powerlaw model has an index $\Gamma$=0.86$\pm$0.23. ULXs are typically observed to have spectral indices of $\Gamma\sim1.9$ \citep{Swartzetal04}, with only a very small fraction having indices as flat as source 3. The source is variable at $>$3$\sigma$ significance, and is located $<$1\arcs\ from the optical centroid of the galaxy. The source luminosity is $L_{0.5-7}$=[9.04$^{+0.66}_{-1.13}$]$\times$10$^{39}$\ergps, and it therefore seems likely that this is the previously undetected active nucleus of NGC~838.

Source 16 is located in the tidal arm extending east from NGC~835. However, the X--ray centroid matches the position of SDSS~J020927.57-100746.2, a faint ($g$=21.8) galaxy with no measured redshift but which is probably unassociated with the group. The source was detected in previous \chandra\ observations \citep[CXO~J020927.6-100746,][]{Evansetal10}. The spectrum is noisy, and cannot distinguish between a $\sim$4~keV thermal plasma or a $\Gamma\sim$1.8 powerlaw. We conclude that this source is probably a background AGN unrelated to HCG~16.

Source 10 is located in the tidally disturbed structure on the southern edge of NGC~835 and NGC~833 and is bright and stable enough to have been previously detected as source CXO J020922.8-100824 \citep{Evansetal10}. The spectra are not sufficient to distinguish between a thermal or powerlaw origin for the emission, but it seems plausible that the source is either a ULX in the tidal zone, or a background AGN. 

Of the fainter sources, three (4, 5 and 13) are bright enough to be ULXs if they are truly point sources located in HCG~16. Source 5 corresponds to another peak in diffuse emission east of the core of NGC~838, and thus may be a gas clump like source 6. Source 4 is at the western edge of the diffuse emission of NGC~838 and is clearly visible in the 2-7~keV band, suggesting that it is a spectrally hard source. Source 13 is located just southwest of the core of NGC~833, close to the position of some unclassified optical sources visible in HST WFPC2 $V$-band imaging. Source 11 is located $<$0.5\arcs\ from SDSS J020938.10-100819.4 a disk galaxy north of NGC~838 with no measured spectroscopic redshift, and a photometric redshift of $\sim$0.2-0.25. 

Based on the cumulative luminosity function of background sources measured by the \chandra\ Multiwavelength Project \citep[ChaMP,][]{Kimetal07a} we are able to estimate the number of background AGN we would expect to find within the \Dtf\ ellipses and tidal interaction regions of the four major galaxies. The \Dtf\ ellipses enclose an area of $\sim$3.05 square arcminutes, with the tidal regions adding $\sim$10\% to this value. Based on the sensitivity limit of the stacked 2013 observations, we would expect to see 2.4 background AGN within the area of the galaxies in the 0.5-8~keV band. Our true sensitivity is probably somewhat better than this, since we include the two earlier observations when detecting sources, but the number of expected sources will also be somewhat lower since we use a narrower energy band (0.5-7~keV) than the ChaMP team. However, since both effects will be relatively small, we still expect to find 2-3 background sources in our regions of interest. This agrees well with our findings that sources 7 and 11 are probably associated with background galaxies.

\section{Galaxy and AGN emission}
\label{sec:AGNdiff}

To examine the radio emission from each galaxy, we extracted total flux densities in each band from regions defined based on the 3$\sigma$ detection contours. These 610~MHz and 1.4~GHz flux densities are listed in Table~\ref{tab:Rflux}. From these fluxes we calculate a spectral index $\alpha$ for each galaxy, defined as $S_\nu\propto\nu^{-\alpha}$, $S_\nu$ being the flux density at frequency $\nu$. We also calculated the radio power at each frequency, where the power $P_\nu$ is defined as $P_\nu$=4$\pi$D$^2$(1+$z$)$^{\alpha-1}S_\nu$, where D is the distance and $z$ the redshift of the group. The 610~MHz data have sufficient spatial resolution to allow us to separate core and diffuse components of the radio emission for NGC~835 and NGC~838, and to suggest that some of the extended emission in NGC~848 may be associated with a background galaxy. We therefore report separate 610~MHz flux densities for the different components of these galaxies. 

\begin{deluxetable*}{llccccc}
\tablewidth{0pt}
\tablecaption{\label{tab:Rflux}Radio fluxes, spectral indices and powers for the five major galaxies}
\tablehead{
\colhead{Galaxy} & \colhead{} & \colhead{$S_{610}$} & \colhead{$S_{1400}$} & \colhead{$\alpha$} & \colhead{$P_{610}$} & \colhead{$P_{1400}$}  \\
\colhead{} & \colhead{} & \colhead{(mJy)} & \colhead{(mJy)} & \colhead{} & \colhead{(10$^{21}$~W~Hz$^{-1}$)} & \colhead{(10$^{21}$~W~Hz$^{-1}$)}\\
}
\startdata
NGC~833 & total & 6.0$\pm$0.3 & 4.1$\pm$0.2 & 0.5$\pm$0.1 & 2.3$\pm$0.1 & 1.6$\pm$0.1 \\
NGC~835 & total & 90.9$\pm$4.6 & 47.6$\pm$2.4 & 0.8$\pm$0.1 & 34.6$\pm$1.8 & 18.1$\pm$0.9 \\
        & core & 75.1$\pm$3.8 & - & - & - & - \\
        & extended & 15.8$\pm$0.8 & - & - & - & - \\
NGC~838 & total & 175.2$\pm$8.8 & 93.1$\pm$4.7 & 0.8$\pm$0.1 & 66.7$\pm$3.4 & 35.5$\pm$1.8 \\
        & core & 151.7$\pm$7.6 & - & - & - & - \\
        & extended & 23.5$\pm$1.2 & - & - & - & - \\
NGC~839 & total & 56.3$\pm$2.8 & 37.5$\pm$1.9 & 0.5$\pm$0.1 & 21.4$\pm$1.1 & 14.2$\pm$0.7 \\
NGC~848 & total & 18.6$\pm$0.9 & 10.9$\pm$0.5 & 0.6$\pm$0.1 & 7.1$\pm$0.3 & 4.1$\pm$0.2 \\
        & no extended & 12.0$\pm$0.6 & - & - & - & -  
\enddata
\end{deluxetable*}

Moving on to the X-ray properties of the major galaxies, we initially extracted spectra from relatively large regions around each of them. Point sources within these regions, except those corresponding to the nuclear sources in NGC~833, NGC~835 and NGC~839, were not excluded, so as to allow an accurate measurement of the total hard X-ray flux from the stellar populations of the galaxies. With the exception of NGC~848, where the shallow off-axis observation contains only a small number of counts, we determined the size of the extraction regions by examining the cumulative growth in the number of counts with radius from the centre of each galaxy, and selected regions which contain 95\% of the background subtracted flux. In the case of NGC~833 and 835 there is a clear region of diffuse emission between the two galaxies, and in the eastern tidal arm extending from NGC~835. We therefore used 15\arcs\ radius circles to extract spectra of the AGN and galaxy cores, and a larger polygonal region to examine the diffuse component.

The galaxy spectra were fitted with models consisting of APEC thermal
plasma and powerlaw components, folded through an absorber whose column was
fixed at the Galactic value. In the three galaxies with bright nuclear
sources, a model of the galaxy emission with the AGN region (as defined in
Table~\ref{tab:sources}) removed was initially fitted. The AGN region was
then reintroduced, and the galaxy model allowed to vary only in overall
normalization, to account for the increased emission volume. Additional
components were then added to model any AGN.  Previous studies suggest that
all three galaxies host partially absorbed AGN, so these were modelled
using a powerlaw observed through a partial-covering absorber at the
redshift of the galaxy. Although the partial covering fraction was allowed
to fit in each case, we found that the best fit value was always 1.0. This
suggests that all three AGN are in fact fully covered by their absorbers,
and any unabsorbed powerlaw component probably arises from the X--ray
binary population in the host galaxy. We outline the results of the fits to
each galaxy below and best-fitting model parameters are shown in
Table~\ref{tab:AGN}.

\begin{deluxetable*}{llcccccccc}
\tablewidth{0pt}
\tablecaption{\label{tab:AGN}Best-fitting spectral model parameters for the five major galaxies}
\tablehead{
\colhead{Component} & \colhead{Parameter} & \colhead{NGC~833} & \colhead{NGC~835} & \multicolumn{2}{c}{NGC~838} & \multicolumn{2}{c}{NGC~839} & \colhead{NGC~848} 
}
\startdata
Model & & AP+PL+abs.PL & AP+PL+abs.PL & AP+PL & AP+AP & AP+PL+abs.PL & AP+abs.PL & PL\\
\hline
\multicolumn{10}{l}{\textit{Galaxy emission}}\\
Soft     & kT$^a$        & 0.47$^{+0.14}_{-0.10}$ & 0.61$\pm$0.03 & 0.80$\pm$0.02          & 0.78$\pm$0.02          & 0.88$\pm$0.05 & 0.79$^{+0.03}_{_-0.04}$ & - \\[+.5mm]
Thermal  & Z$^b$         & 0.47$^{+0.47}_{-0.37}$ & 0.22$^{+0.15}_{-0.07}$ & 0.16$^{+0.08}_{-0.04}$ & 0.16$^{+0.05}_{-0.03}$ & 0.3$^f$ & 0.3$^f$ & - \\[+.5mm]
         & L$^c_{0.5-7}$ & 41.25$^{+8.02}_{-6.88}$ & 200.91$^{+17.95}_{-17.57}$ & 415.56$^{+51.56}_{-64.17}$& 420.15$^{+45.45}_{-37.43}$& 92.81$\pm$11.84 & 106.56$^{+12.60}_{-13.75}$ & - \\[+.5mm]
Hard     & \NH$^d$ & 0.0$^f$ & 0.0$^f$ & 0.0$^f$ & 0.0$^f$ & 0.0$^f$ & 0.47$^{+0.24}_{-0.18}$ & 0.0$^f$\\[+0.5mm]
         & $\Gamma$      & 1.65$^f$               & 1.65$^f$               & 1.84$^{+0.18}_{-0.19}$ & kT$^a$=4.27$^{+2.28}_{-0.85}$ & 1.46$^{+0.15}_{-0.16}$ & 1.67$^{+0.17}_{-0.15}$ & 1.65$^{+0.28}_{-0.25}$ \\[+.5mm]
         & L$^c_{0.5-7}$ & 79.83$^{+10.31}_{-11.84}$ & 76.01$^{+30.56}_{-32.08}$ & 409.83$^{+59.58}_{-56.53}$& 376.99$^{+36.67}_{-34.38}$ & 207.78$\pm$17.57 & 477.44$^{+48.89}_{-35.14}$ & 212.36$^{+47.36}_{-42.78}$ \\[+.5mm]
         & red. $\chi^2$/d.o.f$^e$ & 1.54/23         & 1.10/48               & 1.12/233             & 1.10/233             & 0.98/73 & 0.99/109 & 0.59/5 \\
\hline
\multicolumn{7}{l}{\textit{Nuclear emission}}\\
AGN      & \NH$^d$       & 22.42$^{+7.84}_{-6.30}$& 23.36$^{+3.74}_{-3.23}$& - & - & 0.92$^{+0.99}_{-0.55}$ & - & - \\[+.5mm]
         & $\Gamma$      & 0.46$^{+0.96}_{-0.82}$ & 0.67$^{+0.48}_{-0.42}$ & - & - & 1.85$^{+0.55}_{-0.37}$ & - & - \\[+.5mm]
         & L$^c_{0.5-7}$ & 1519.4$^{+1551.5}_{-467.5}$& 6924.8$^{+2979.2}_{-41.3}$& - & - & 274.62$^{+141.70}_{-46.98}$ & - & - \\[+.5mm]
Fe K$\alpha$ & Energy$^a$ & - & 6.41$\pm$0.05 & - & - & - & - & - \\[+.5mm]
         & width $\sigma$ & - & 0.10$\pm$0.05 & - & - & - & - & - \\[+.5mm]
         & L$^c_{0.5-7}$ & - & 116.11$^{+45.45}_{-40.87}$ & - & - & - & - & - \\[+.5mm]
         & red. $\chi^2$/d.o.f$^e$ & 0.85/56 &  1.09/174 & - & - & 1.04/110 & - & -
\enddata
\tablecomments{Model indicates major model component, APEC thermal plasma (AP), powerlaw (PL) and powerlaw with intrinsic absorption (abs.PL). $^a$ Temperatures and energies are in units of keV. $^b$ Abundance relative to the solar value. $^c$ 0.5-7~keV luminositieses are given in units of 10$^{38}$\ergps. $^d$ Intrinsic absorption in units of 10$^{22}$\pcmsq. $^e$ The reduced $\chi^2$ of the fit over its degrees of freedom. $^f$ Parameters fixed during fitting.}
\end{deluxetable*}

\subsection{NGC~833 / HCG~16B}
NGC~833 is a disturbed Sb galaxy, with an asymmetric velocity distribution and misalignment between the gas and stellar rotation axes \citep{MendesdeOliveiraetal98}, indicating that it is interacting with NGC~835. The galaxy has a LINER nucleus \citep{VeronCettyVeron06} but there is no optical evidence of ongoing star formation, and the galaxy is relatively poor in ionized and molecular gas compared to its companions \citep{MendesdeOliveiraetal98}. It is also the least luminous of the four major galaxies in the infrared and ultraviolet bands. Using the \xmms\ first light observation, \citet{Turneretal01special} find that the galaxy spectrum is best modelled using a combination of a soft (0.47$\pm$0.12~keV) thermal model and two powerlaw components with $\Gamma$=1.8$\pm$0.5, one of which is strongly absorbed below $\sim$3~keV.

\begin{figure}
\includegraphics[width=\columnwidth,bb=36 126 576 666]{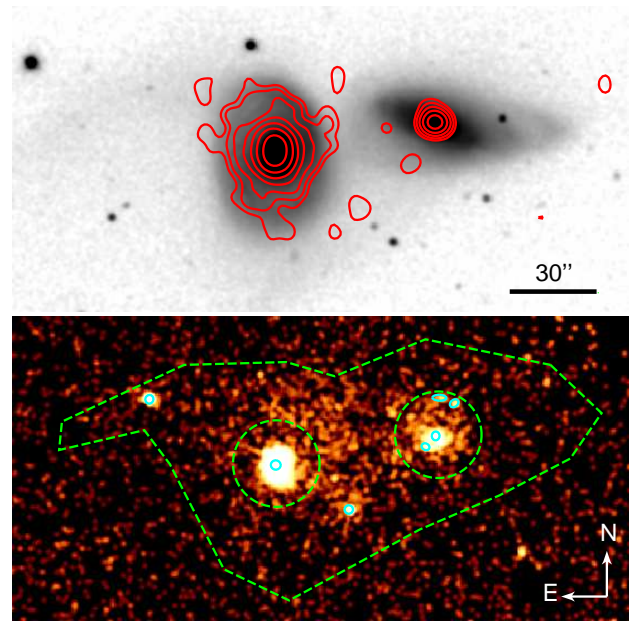}
\caption{\label{fig:sourcesAB} Images of NGC~835 (left) and NGC~833
  (right). The upper panel shows an SDSS $i$-band image with contours of
  GMRT 610~MHz flux density overlaid (starting at 3$\times$rms and
  increasing in steps of factor 2). The lower panel shows a \chandra\ 0.5-2
  keV image with the same alignment and scale, smoothed with a 2\arcs\
  Gaussian. Small cyan ellipses indicate spectral extraction regions used
  to examine point sources. Dashed regions are those used to extract
  spectra of the galaxy cores and associated diffuse emission. }
\end{figure}

\begin{figure}
\includegraphics[width=\columnwidth,bb=10 30 710 585,clip=]{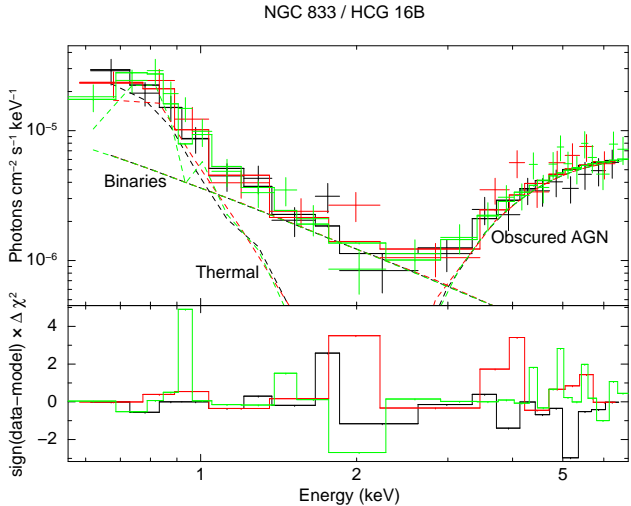}
\caption{\label{fig:HCG16b_spec} Spectra and best fitting model for the
  central 15\arcs-radius region of HCG~16B / NGC~833. Green, black and red
  points represent ObsIDs 15181, 15666 and 15667 respectively, solid lines
  show the best fitting model and dashed lines contributions from the
  various model components. The lower panel shows the significance of the
  residuals to the fit.}
\end{figure}

The GMRT 610~MHz image shows only an unresolved source at the position of
NGC~833, presumably associated with the AGN. We model the X-ray spectrum of
the central 15\arcs-radius region of the galaxy with the thermal plus
powerlaw plus absorbed powerlaw model described above. The spectral
extraction region, and those of point sources in the galaxy, is shown in
Figure~\ref{fig:sourcesAB}. Our best fit to spectra from ObsIDs 15181,
15666 and 15667 is shown in Figure~\ref{fig:HCG16b_spec} and its parameters
are listed in Table~\ref{tab:AGN}. The temperature of the thermal component
agrees with that found by TRP01, and we find a roughly half solar
abundance. Our absorbed AGN model has a hydrogen column identical within
errors with that found by TRP01, and a flatter powerlaw index consistent
with TRP01 at the 2$\sigma$ level.

NGC~833 is also in the field of view of ObsID~923. The observation is too short to constrain our chosen model. However, comparing our best fitting model to the spectrum we find that the soft emission is reasonably well described by the thermal component, but that the hard emission is stronger than the model predicts. To test whether this arises from a change in the AGN intrinsic luminosity or in the absorbing column, we allow either the powerlaw normalization or the column density to vary while holding all other parameters fixed at their best fit values. An increase in normalization provides a significantly better fit than an increase in absorbing column (reduced $\chi^2$=1.457 compared to 2.772 for 10 degrees of freedom), leading us to conclude that in 2000 November the AGN luminosity was a factor $\sim$3 greater than in 2013 July.

\subsection{NGC~835 / HCG~16A}
The other galaxy in the interacting pair, NGC~835, is a Seyfert~2 \citep{VeronCettyVeron06} with an apparent tidal tail and clumpy ring of UV-bright knots at the edge of its disk, suggestive of star formation (TRP01). As with NGC~833, TRP01 modelled the X-ray emission from the galaxy with a combination of thermal and absorbed powerlaw components, finding that the temperature of the thermal component fell from $\sim$0.5~keV in the galaxy core to $\sim$0.3~keV in the star forming ring. The GMRT 610~MHz data reveal a central point source surrounded by diffuse emission from the galaxy disk, probably arising from star formation.

\begin{figure}
\includegraphics[width=\columnwidth,bb=10 30 710 585,clip=]{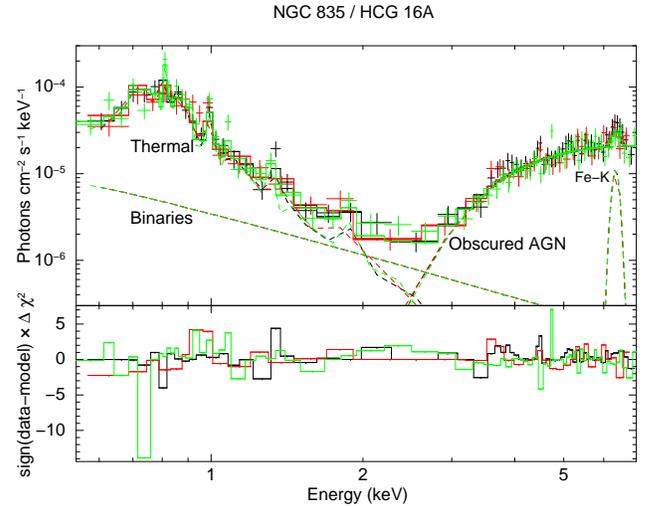}
\caption{\label{fig:HCG16a_spec} Spectra and best fitting model for the central 15\arcs-radius region of HCG~16A / NGC~835, using ObsIDs 15181, 15666 and 15667. Colors and lines are as described in Figure~\ref{fig:HCG16b_spec}.}
\end{figure}

We extract spectra from a 15\arcs-radius region (inside the ring) and fit the same model we used for NGC~833. The spectral extraction region is shown in Figure~\ref{fig:sourcesAB}. Fitting the 2013 observations simultaneously, we find a satisfactory solution, with an apparent 6.4~keV Fe K$\alpha$ line visible in the longest exposures, ObsIDs 15181 and 15667. We therefore add a redshifted Gaussian component to the model and allow its energy, flux and line width to fit. The best fitting parameters are shown in Table~\ref{tab:AGN} and the best fitting model in Figure~\ref{fig:HCG16a_spec}. 

Our best fitting thermal component has a temperature higher than, but consistent at the 2$\sigma$ level with, that found by TRP01. We find an an abundance 0.25\Zsol\ lower than that in NGC~833, but consistent within the uncertainties. The AGN component has a significantly lower absorbing column than that found from \xmms\ by TRP01 (23.4$^{+3.7}_{-3.2}$$\times$10$^{22}$\pcmsq\ compared to 46$\pm$15$\times$10$^{22}$\pcmsq) and a flatter powerlaw index (0.67$^{+0.48}_{-0.42}$ compared to 2.25$\pm$0.23). However, we note that our model differs from that of TRP01; their model contains no component to account for the X-ray binary population, but does include an unabsorbed AGN component whose powerlaw index is partially constrained by soft emission.

As with NGC~833, we find that our best fitting thermal component provides a reasonable description of the spectra extracted from ObsIDs 923 and 10394 below $\sim$2~keV, but that the hard 2-7~keV band flux in these earlier observations is significantly less than expected. We again allow either the AGN powerlaw normalization or absorbing column to fit, to test the likely cause of the reduced hard fluxes. In both cases a reduced normalization produces the better fit; reduced $\chi^2$=1.374 compared to 1.559 for 13 degrees of freedom in ObsID 923 and reduced $\chi^2$=2.098 compared to 2.559 for 12 degrees of freedom in ObsID 10394. The best fitting normalisations are $\sim$21\% and $\sim$18\% of the value found for the 2013 observations, respectively. However, the poor quality of these fits indicates that these simple adjustments to the model are insufficient, and the overall shape of the spectrum may have changed. Unfortunately these earlier observations lack the depth to constrain the model further. We conclude that the Seyfert nucleus of NGC~835 is variable in the X-ray band on timescales of months to years, probably in large part owing to changes in intrinsic luminosity.

\subsection{Diffuse emission between NGC~833 and NGC~835}
We extract spectra of the diffuse emission around and between NGC~833 and NGC~835 using a polygon region chosen to follow the stellar envelope, excluding point sources and the 15\arcs\ radius galaxy regions described above (see Figure~\ref{fig:sourcesAB}. The region includes the tidal arm and star--forming ring of NGC~835, and the outer tidally-disturbed parts of NGC~833. All five observations cover this region, though only partially in the case of ObsID~10394. The spectra are adequately modelled by a thermal plasma model with Galactic absorption (reduced $\chi^2$=1.06 for 108 degrees of freedom). The thermal component has temperature kT=0.48$^{+0.05}_{-0.04}$~keV, abundance 0.09$^{+0.04}_{-0.03}$\Zsol\ and a 0.5-7~keV luminosity of (1.08$\pm$0.05)$\times$10$^{40}$\ergps. Neither a two-temperature thermal model or a thermal plus powerlaw model provide a better fit, and both have unconstrained parameters, suggesting that the diffuse emission is at least primarily thermal and relatively soft.

\subsection{NGC~838 / HCG~16C}
NGC~838 is a Luminous Infrared Galaxy (LIRG) having an infrared luminosity
$>$10$^{11}$\Lsol, and is the most IR and UV luminous of the four major
galaxies. \citet{MendesdeOliveiraetal98} note a number of optical features
indicating disturbance and possibly a recent merger, including kinematic
warping and multiple kinematic components in the ionized gas which are
misaligned with the stellar major axis. \citet{Vogtetal13} argue that the
galaxy underwent a period of global star formation $\sim$500~Myr ago, with
star formation continuing in the galaxy core. This activity is driving an
asymmetric bipolar wind, visible in ionized gas, which is probably only a
few Myr old in its current phase \citep{Vogtetal13}. TRP01, using \xmms,
identified diffuse X-ray emission in the galaxy, and found that it could be
modelled as two-temperature thermal emission with kT=0.59$\pm$0.04~keV and
3.2$\pm$0.8~keV.

\begin{figure}
\includegraphics[width=\columnwidth,bb=36 243 576 548]{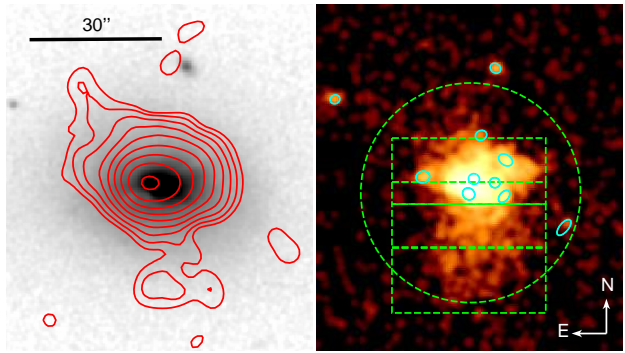}
\caption{\label{fig:sourcesC}Images of NGC~838. The left panel shows an SDSS $i$-band image with contours of GMRT 610~MHz flux density overlaid (starting at 3$\times$rms and increasing in steps of factor 2). The right panel shows a \chandra\ 0.5-2 keV image with the same alignment and scale, smoothed with a 2\arcs\ Gaussian. Small cyan ellipses indicate spectral extraction regions used to examine point sources. Dashed regions are those used to extract spectra of the galaxy as a whole and its superwind.
}
\end{figure}

The \chandra\ image (Figure~\ref{fig:sourcesC}) shows that the diffuse
emission extends along a roughly north-south axis to $\sim$15\arcs\ north
of the galaxy centroid and $\sim$25\arcs\ southward. The northern emission
appears brighter. \citet{Vogtetal13} suggest that the northern side of the
galaxy disk is facing us with the central starburst obscured by a dust
lane, and that while the outflowing starburst wind has inflated bubbles on
both sides of the disk, the northern bubble is better confined by the
surrounding neutral gas and therefore more compact. The \chandra\ X-ray
images are consistent with this picture, and we also see 610~MHz emission
extending $\sim$30\arcs\ north and south of the galactic disk. In the south
this is clearly correlated with the X-ray (and H$\alpha$) emission, but on
the north side of the galaxy the radio source extends furthest to the
northeast and is less clearly correlated with the X-ray emission. The
connection between the starburst winds and the surrounding IGM is discussed
in paper II.

\begin{figure}
\includegraphics[width=\columnwidth,bb=10 30 710 585,clip=]{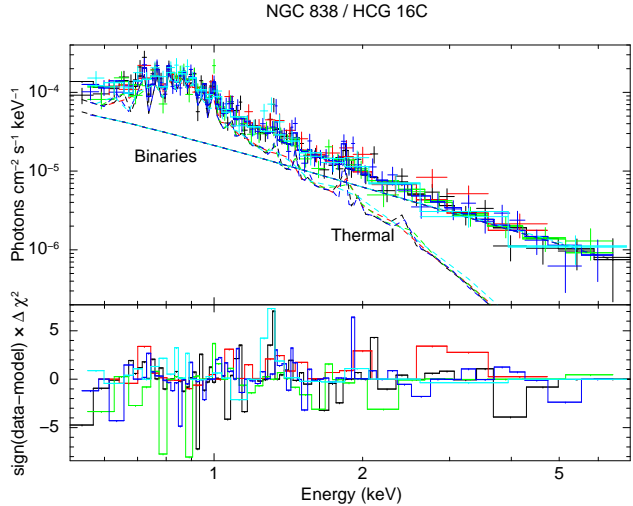}
\caption{\label{fig:HCG16c_spec} Spectra and best fitting APEC+powerlaw model for the central 25\arcs-radius region of HCG~16C / NGC~838, showing all five \chandra\ observations. Colors and lines are as described in Figure~\ref{fig:HCG16b_spec} with ObsID~923 marked in blue and ObsID~10394 in cyan.}
\end{figure}

Extracting spectra from a $\sim$25\arcs\ radius circular region, we find that all five ObsIDs can be adequately modelled by either a two-temperature thermal plasma, or a low-temperature thermal model plus a powerlaw. As expected, there is no indication of significant luminosity variation with time. Best-fitting parameters for both models are listed in Table~\ref{tab:AGN} and APEC plus powerlaw model fit to all five spectra is shown in Figure~\ref{fig:HCG16c_spec}. We find a temperature for the soft component $\sim$0.2~keV hotter than the TRP01 estimate ($\sim$4.5$\sigma$ significant), and a low abundance, Z$\sim$0.16\Zsol. While we might expect a high metallicity in a galactic wind driven by supernovae, it will be diluted by entrainment and mixing with less enriched cold gas. Our measurements may be biased low by the multiphase nature the wind, since a single-temperature model will typically underestimate the abundance of multi-temperature gas around 1~keV \citep[the ``Fe bias'',][]{Buotefabian98,Buote00b}. As suggested by TRP01, the spectrally hard component probably arises from young high-mass X-ray binaries formed in the star formation episodes, while the soft component is primarily the hot gas of the starburst wind. 

To examine the temperature structure of the wind we divided the emission
into strips with widths of 5, 10, and 15\arcs\ (chosen to include a few
hundred counts) and extracted spectra from all five observations. A 5\arcs\
width strip was placed at the optical centroid of the galaxy, and includes
the majority of the hard (2-7~keV) emission. Figure~\ref{fig:sourcesC}
shows these regions. We fitted each region with an APEC or (in the central
region) APEC+powerlaw model, either allowing abundance to fit freely or
fixing it at 0.3\Zsol. The results of these fits for the thermal component
are shown in Table~\ref{tab:wind}. Although the models are
under-constrained in some cases (reduced $\chi^2<1$), the fits suggest that
the wind has a fairly consistent temperature of $\sim$0.7-0.8~keV, falling
to 0.3~keV (a 4.4$\sigma$ significant decline) in the outer part of the more
extended southern outflow.

\begin{deluxetable}{lccccc}
\tablewidth{0pt}
\tablecaption{\label{tab:wind}Best-fitting model parameters for the NGC~838 wind}
\tablehead{
\colhead{Region} & \multicolumn{3}{c}{Abundance free} & \multicolumn{2}{c}{Abundance=0.3\Zsol} \\
\colhead{} & \colhead{kT} & \colhead{Abund.} & \colhead{red. $\chi^2$/dof} & \colhead{kT} &  \colhead{red. $\chi^2$/d.o.f.} \\
\colhead{} & \colhead{(keV)} & \colhead{(\Zsol)} & \colhead{} & \colhead{(keV)} & \colhead{} \\
}
\startdata
Central & 0.92$^{+0.05}_{-0.04}$ & $>$0.14 & 1.28/101 & 0.92$^{+0.05}_{-0.04}$ & 1.26/102 \\
North 1 & 0.79$\pm$0.02 & 0.08$\pm$0.01 & 1.34/92 & 0.81$\pm$0.02 & 2.41/93 \\
South 1 & 0.74$^{+0.06}_{-0.07}$ & 0.06$\pm$0.03 & 0.67/23 & 0.81$\pm$0.04 & 1.47/24 \\
South 2 & 0.30$^{+0.07}_{-0.03}$ & $<$0.03 & 0.57/9 & 0.75$^{+0.08}_{-0.35}$ & 1.36/10
\enddata
\end{deluxetable}

The APEC thermal plasma model assumes collisional ionisation equilibrium, which may not hold in a complex, rapidly expanding galaxy wind \citep[e.g.,][]{Breitschwerdt03}. We therefore test the impact of using a non-equilibrium ionisation (NEI) model instead of APEC in our fits. In general the NEI model produces similar temperatures, abundances and fit statistics. Where significant differences are found (e.g., in the North 1 bin where the best fit NEI temperature is 0.70$\pm$0.02~keV) the NEI model produces a poorer fit to the data (reduced $\chi^2$=1.479). We therefore conclude that, while the wind gas is likely to be out of collisional equilibrium in some regions, our results are unlikely to be significantly effected. Deeper observations would be required to probe the equilibrium state of the wind.

\subsection{NGC~839 / HCG~16D}
NGC~839 is the second LIRG in HCG~16, and has also been classified as a
LINER-2 \citep{DeCarvalhoCoziol99}, though deep optical integral field
spectroscopy suggests that the LINER emission arises from shock excitation
in an outflowing starburst wind \citep{Richetal10}. This wind is visible as
a biconical polar outflow in H$\alpha$ emission, and stellar population
modelling shows that the galaxy contains a sizable population of A stars,
indicating a starburst age of $\sim$400~Myr. TRP01 show that the
\xmms\ spectrum is best fit by a model including a fairly heavily absorbed
powerlaw, and conclude that an active nucleus is present. Our GMRT 610~MHz
observation detects only an unresolved source coincident with the galaxy
centroid.

\begin{figure}
\includegraphics[width=\columnwidth,bb=36 274 576 517]{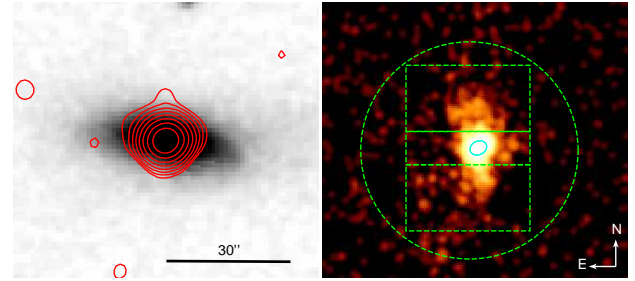}
\caption{\label{fig:sourcesD}Images of NGC~839. The left panel shows an Digitized Sky Survey $R$-band image with contours of GMRT 610~MHz flux density overlaid (starting at 3$\times$rms and increasing in steps of factor 2). The right panel shows a \chandra\ 0.5-2 keV image with the same alignment and scale, smoothed with a 2\arcs\ Gaussian. Small cyan ellipses indicate spectral extraction regions used to examine point sources. Dashed regions are those used to extract spectra of the galaxy as a whole and its superwind.}
\end{figure}

The galaxy is outside the field of view of ObsID~10394, but we extract
spectra from the other four observations, using the regions shown in
Figure~\ref{fig:sourcesD}. We initially extract a spectrum excluding the
central point source, and find that this is well fitted by an APEC plus
powerlaw model, representing the thermal emission from the galactic wind,
and the X-ray binary population. Applying this model to the spectra for the
whole galaxy, we find that the emission from the central point source is
best modelled by a second, intrinsically absorbed powerlaw. 

\begin{figure}
\includegraphics[width=\columnwidth,bb=10 30 710 585,clip=]{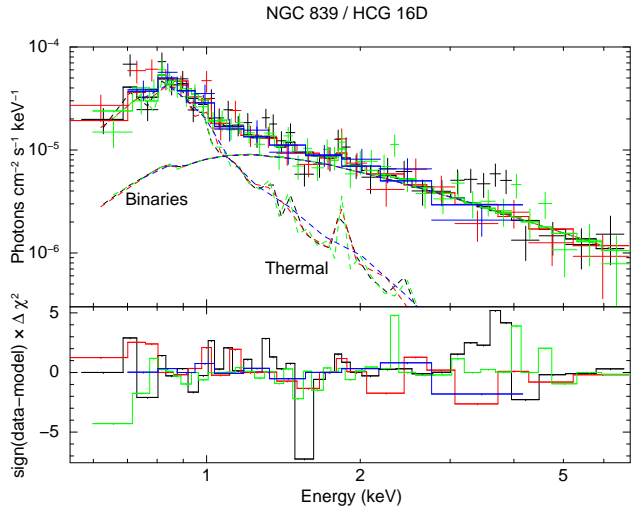}
\caption{\label{fig:HCG16d_spec} Spectra and best fitting APEC+absorbed powerlaw model for the central 26\arcs-radius region of HCG~16D / NGC~839, showing data from ObsIDs 923, 15181, 15666 and 15667. Colors and lines are as described in Figure~\ref{fig:HCG16b_spec} with ObsID 923 marked in blue.}
\end{figure}

However, we find that the emission from the galaxy as a whole (including
the central point source) can also be well modelled using an APEC component
and a single intrinsically-absorbed powerlaw. This model fit is shown in
Figure~\ref{fig:HCG16d_spec}. The fact that the hard component of the
spectra can be modelled by a single absorbed powerlaw component suggests
the possibility that it could be dominated by emission from an AGN, or that
the X-ray binary population dominates and the AGN emission is negligible.
In either case, the model requires a significant absorption column within
the galaxy. Parameters for both fits are listed in Table~\ref{tab:AGN}. The
abundance of the thermal component is poorly constrained and is therefore
fixed at 0.3\Zsol\ in both fits.

Our best fitting temperature is somewhat higher than that found by TRP01 (0.88$\pm$0.05 or 0.79$^{+0.03}_{-0.04}$ compared to 0.63$\pm$0.10). The powerlaw index is consistent with the value found by TRP01 (1.85$^{+0.55}_{-0.37}$ or 1.67$^{+0.17}_{-0.15}$ compared to 2.1$\pm$0.8) but we find a much lower absorbing column, [0.92$^{+0.99}_{-0.55}$] or [0.47$^{+0.24}_{-0.18}$]$\times$10$^{22}$\pcmsq\ compared to their best fitting value of [45$\pm$20]$\times$10$^{22}$\pcmsq. We will discuss the origin of the powerlaw emission, and the question of whether NGC~839 hosts an AGN, in Section~\ref{sec:disc839}

As with NGC~838, there is enough diffuse emission from the wind regions north and south of the galaxy core to allow (crude) spectral fitting. We divide the galaxy into three regions (8 and 16\arcs\ wide strips), fitting the core with the model described above, and the wind regions with an absorbed APEC thermal plasma model. The wind emission is too faint to constrain abundance and we therefore fix it at 0.3\Zsol. The parameters of the thermal components of the fits are listed in Table~\ref{tab:hcg16dwind}.

\begin{deluxetable}{lcc}
\tablewidth{0pt}
\tablecaption{\label{tab:hcg16dwind}Best-fitting model parameters for the NGC~839 wind}
\tablehead{
\colhead{Region} & \colhead{kT} & \colhead{red. $\chi^2$/d.o.f.} \\
\colhead{} & \colhead{(keV)} & \colhead{} \\
}
\startdata
Central & 0.92$^{+0.05}_{-0.10}$ & 0.922/74 \\[+0.5mm]
North & 1.24$^{+0.11}_{-0.12}$ & 1.652/8 \\[+0.5mm]
South & 0.94$^{+0.09}_{-0.12}$ & 0.701/7 \\[+0.5mm]
North+South & 0.99$\pm$0.07 & 1.183/17
\enddata
\tablecomments{Abundance fixed at 0.3\Zsol\ in all fits.} 
\end{deluxetable}

\subsection{NGC~848}
\begin{figure}
\includegraphics[width=\columnwidth,bb=36 254 576 538]{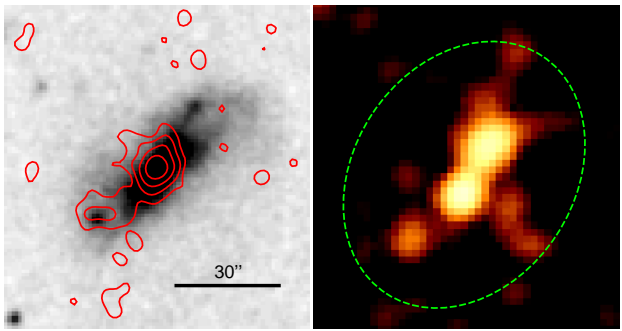}
\caption{\label{im:N848} Images of NGC~848. The left panel shows a \textit{Digitized Sky Survey} $R$-band image with contours of GMRT 610~MHz flux density overlaid (starting at 3$\times$rms and increasing in steps of factor 2). The right panel shows a \chandra\ 0.5-2 keV image with the same alignment and scale, binned to 2\arcs\ pixels and smoothed with a 6\arcs\ Gaussian. The dashed region indicates the \Dtf\ ellipse used to extract a spectrum of the galaxy emission. The two images have the same scale and orientation.}
\end{figure}

NGC~848 lies $\sim$15\arcm\ from the group core, $\sim$11.5\arcm\ southeast
of NGC~839, but is linked to the four main galaxies by the \Hi\ filament \citep{VerdesMontenegroetal01}. It is a peculiar barred spiral galaxy hosting a starburst \citep{Continietal98}, and is only visible on the ACIS-S1 CCD of observation 923. There is a clear detection of emission from the galaxy, apparently extended along the galactic bar (see Figure~\ref{im:N848}) though the large off-axis angle means that individual sources cannot be resolved.

We extract a spectrum from the S1 data using an elliptical source region
corresponding to the \Dtf\ optical contour of the galaxy, and a background
spectrum from an annulus centred on the galaxy with radii 1.75\arcm\ and
3.5\arcm. The effective exposure is short enough that even a simple
absorbed powerlaw model is under-constrained, with reduced $\chi^2$=0.586
for 5 degrees of freedom.  The best fitting parameters of this model are
listed in Table~\ref{tab:AGN}. The powerlaw index,
$\Gamma$=1.65$^{+0.28}_{-0.25}$, is consistent with emission from X-ray
binaries, but is also consistent with emission from a LINER or Seyfert
nucleus \citep[e.g.,][]{GuCao09}. However, comparison with the optical and
radio images shows that the galaxy core is located between the two most
luminous clumps of X-ray emission, indicating that any X-ray emission from
an AGN is weak compared to the emission associated with star formation and
the X-ray binary population.  The spectral model gives luminosity
$L_{0.5-7}$=2.12$^{+0.47}_{-0.43}$$\times$10$^{40}$\ergps\ at our adopted
distance for the group.  An APEC thermal model provides a fit of similar
quality, but the temperature is unphysical (kT=5.1$^{+8.4}_{-2.4}$~keV),
suggesting there is relatively little hot gas in the galaxy. Fixing the
temperature to kT=0.5~keV and the abundance to Z=0.3\Zsol, we can place a
3$\sigma$ upper limit on the flux from any thermal component of
$F_{0.5-7}<$1.38$\times$10$^{-14}$\ergpspcmsq\
($L_{0.5-7}<$5.27$\times$10$^{39}$\ergps), fainter than the thermal
emission from any of the other major group members except NGC~833.  The
GMRT 610~MHz image shows a marginally extended central source, with an
extension to the southeast that appears to correspond to an unidentified
optical source. This may be a foreground star or background AGN.

\section{Stellar Population Modelling}
\label{sec:SSP}
The Sloan Digital Sky Survey partially covers HCG~16, and optical spectra
of the cores of NGC~833 and NGC~838 are available as part of Data Release
10 \citep[SDSS-DR10,][]{Ahnetal14}. We analysed these spectra and
  estimated the ages, metallicities, and mass of the stellar populations
  using the spectral fitting code STARLIGHT \citep{CidFernandesetal05}.
STARLIGHT fits the observed spectrum with a combination of simple stellar
population (SSP) models covering a range of ages and metallicities. The
code returns the contribution, as a fraction of total stellar mass, from
each {\it basis} SSP. Before running the code, the observed spectra are
corrected for foreground extinction and de-redshifted, and the models are
degraded to match the wavelength-dependent resolution of the spectrum of
each galaxy, as described in \citet{LaBarberaetal10}.

In order to check the model-dependency of the stellar population properties we used two sets of SSP models. One of them is based
on the Medium resolution INT Library of Empirical Spectra \citep[MILES,
][]{SanchezBlazquezetal06}, using the updated version 9.1
\citep{FalconBarrosoetal11} of the code presented in
\citet{Vazdekisetal10}. We selected models computed with \citet{Kroupa01}
universal initial mass function (IMF) and isochrones by
\citet{Girardietal00}. We also used \citet[][BC03]{BruzualCharlot03}
models, calculated with Padova 1994 evolutionary tracks
\citep{Girardietal96} and with \citet{Chabrier03} IMF. The basis grids
cover ages in the range $0.07 - 14.2$~Gyr for MILES models and $0.02
-14.2$~Gyr for BC03 models, with constant $\log({\rm Age})$ steps of 0.2.
We selected SSPs with metallicities [M/H]~=$\{-1.71, -0.71, -0.38, 0.00,
+0.20\}$.

For NGC~833, we adopted the \citet{Cardellietal89} extinction law ($R_{\rm
  V} = 3.1$). NGC~838 contains several dust lanes, and we adopted
the \citet{Calzettietal00} law ($R_{\rm V} = 4.05$), which is more suitable for
starburst galaxies. We also allowed stellar populations younger than
$0.032$~Gyr to have an extra extinction in relation to the older
populations. The stellar masses -- computed within the fiber aperture --
are corrected to the full extent of the galaxy by computing the difference
between fiber and model magnitudes in the $z$ band. However, we note that
this correction is only approximate, since the stellar population in the fiber
aperture is probably not representative of the galaxy as a whole.

In NGC~833, we find that the galaxy is dominated by an old stellar
population (mass-weighted age $\sim$10~Gyr for MILES and $\sim$13.5~Gyr for
BC03), with a young component making up $\sim$0.01\% of the stellar mass.
The age of the young component is dependent on the models used, with star
formation beginning $\sim$300~Myr ago for MILES and $\sim$50~Myr ago for
BC03. However, in both cases the star formation rates are small
(1-3\Msolpyr) and continue to the lower age limit of the basis grid. The
fits suggest a total stellar mass for the galaxy of
1-4$\times$10$^{11}$\Msol\ (in rough agreement with luminosity based
estimates, see Section~\ref{sec:gascont}), and metallicity
[M/Fe]=0.02-0.22.

For NGC~838, the properties of the stellar population are much more
sensitive to the choices made in the spectral fitting. Depending on the
models and extinction law adopted, the mass-weighted age varies from
$\sim$5 to $\sim$11~Gyr, metallicity varies across the range [M/Fe]=-1.2 to
[M/Fe=+0.3, and stellar masses derived with BC03 models are a factor
$\sim$10 greater than those derived from MILES models. The MILES fits find
that 30-50\% of the stars in the fiber aperture were formed in a burst
starting $\sim$300~Myr ago, with SFR $\sim$15-50\Msolpyr. The BC03 fits
suggest a much smaller young component ($\sim$2\% by mass) formed in a
burst starting 50~Myr ago and continuing to the present, with SFR
$\sim$50-80\Msolpyr, and an extended star-formation history for the old
stellar component, continuing as late as 2~Gyr ago.

\begin{figure}
\includegraphics[width=\columnwidth,bb=0 0 576 576]{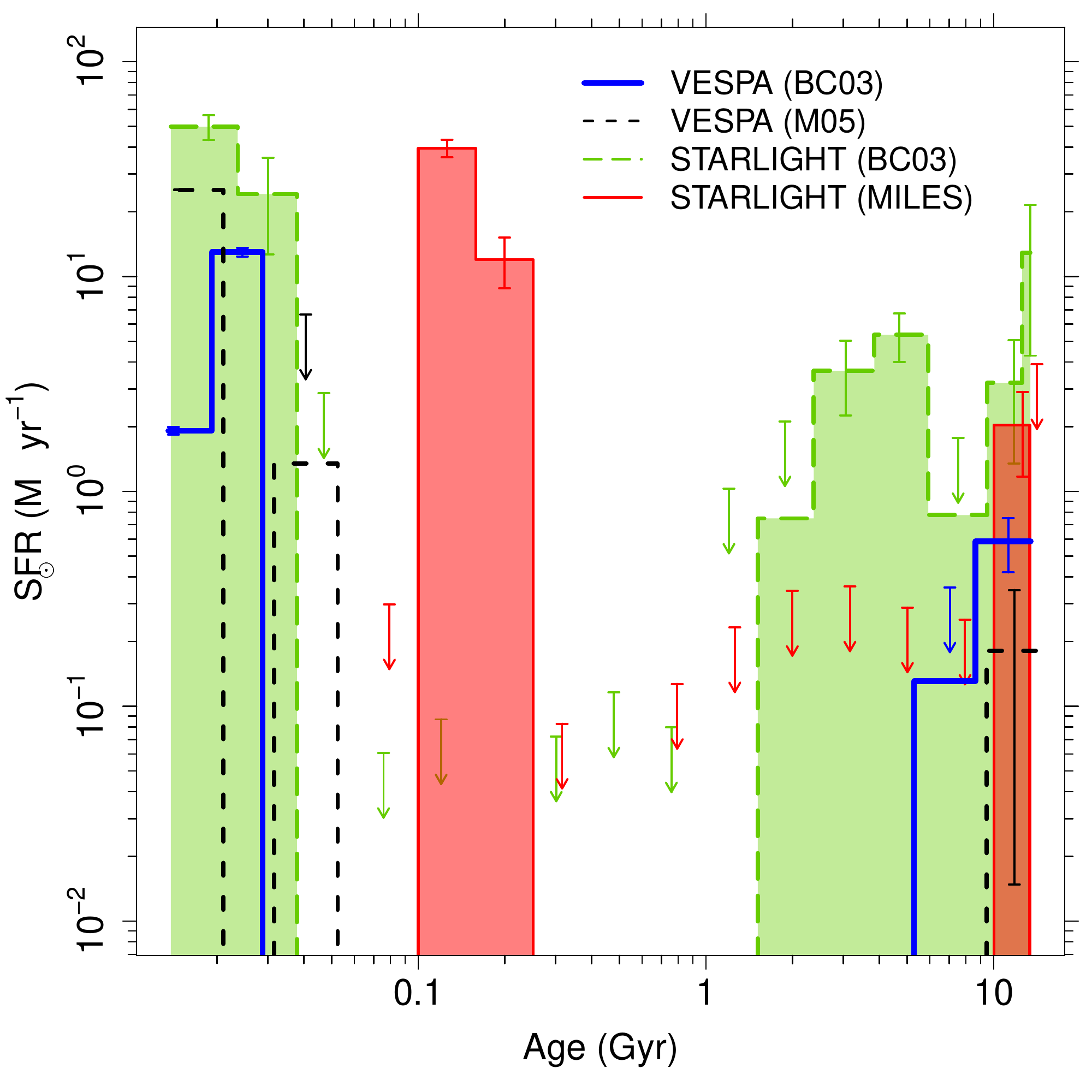}
\caption{\label{fig:SSP838} Star formation history of NGC~838 comparing our fits using the MILES and BC03 models, to results from the VESPA database, using BC03 or \protect\citet{Maraston05} models. Histograms show the results of the model fits, while errorbars and upper limits indicate uncertainties derived from monte-carlo simulations of the best-fitting STARLIGHT model.}
\end{figure}

Figure~\ref{fig:SSP838} shows the results of two fits to the NGC~838
spectrum using the MILES and BC03 models, compared with fits to the same
spectrum from the Versatile Spectral Analysis database
\citep[VESPA,][]{Tojeiroetal09}. VESPA fits use either the BC03 models or
the models of \citet{Maraston05}, and for the young population these
produce similar results to our BC03 fits. However, they differ in finding
little evidence of an extended period of formation for the old population.

To test the reliability of the STARLIGHT fits, we simulated spectra based on the BC03 and MILES modelling results. For each of the the two best-fitting model, 100 spectra were simulated and noise was added to acheive a signal-to-noise ratio of 60, matching that of the observed spectrum. Each simulated spectrum was then fitted using the STARLIGHT code, and the results of the fits used to estimate the 1$\sigma$ uncertainty on the SFR in each age bin, shown as errorbars in Figure~\ref{fig:SSP838}. Where the 1$\sigma$ uncertainty was consistent with zero, show the 1$\sigma$ upper limit on SFR. These uncertainties indicate that the general picture of the star formation history presented by each set of fits is fairly robust, and that even quite low SFRs ($\sim$0.1\Msolpyr) are unlikely in the 300~Myr-1~Gyr age range.   

We must also consider previous observations of NGC~838, which show that the
stellar population in the galaxy core is probably not representative of the
galaxy as a whole. While \citet{Vogtetal13} did not fit SSP models to their
integral field spectra of NGC~838, they did examine those spectra for
simple age indicators. Examining spectra from the core and disk they found
that both regions have similar absorption features typical of a young,
A-type stellar population with an age $\sim$500~Myr post-starburst, but
that the core has a much stronger blue continuum, indicative of an O/B-type
stellar population consistent with very recent/ongoing star formation. They
conclude that the galaxy underwent a galaxy-wide star formation episode
starting $\sim$500~Myr ago, but that star formation rates have declined
with time and activity is now restricted to the core. Our BC03 fits (and
the VESPA results) agree with Vogt et al.  in finding ongoing star
formation in the core, and suggest that it began $\sim$50~Myr ago. Our MILES
fit suggests that star formation began in the core as early as
$\sim$300~Myr ago, but does not find it to be ongoing. However, since the
SDSS spectrum only samples the core, we cannot compare our age estimates
with the overall burst age estimated by Vogt et al. based on spectra from
the disk of the galaxy. We therefore adopt their burst age of
$\sim$500~Myr, with the caveat that star formation probably began more
recently in the galaxy core. High-resolution optical spectra from elsewhere
in NGC~838 would be required to more accurately measure the burst age.

\section{Discussion}

\subsection{Point Sources}
\citet{Tzanavarisetal14} have used ObsID 923 to examine the point source population in and around the four original group members. Eight of their sources are located within our adopted \Dtf\ regions, including the cores of NGC~833, NGC~835 and NGC~839, and several sources in NGC~838. In general our results agree well with theirs for the brighter sources, though our ability to model the spectra of the absorbed AGN naturally produces more accurate powerlaw spectral indices. We find somewhat lower luminosities for the faintest sources, and fail to detect their source 46, probably owing to the different flux extraction methods and background regions used.

\subsection{Gas content of the galaxies}
\label{sec:gascont}
Table~\ref{tab:mgas} shows our estimates of the density and mass of the gas
in the four main galaxies. We estimate the electron number density of the
hot gas ($n_e$) from the spectral fits described in section~\ref{sec:AGNdiff}.
The core region of NGC~833 is reasonably well described by a spherical
$\beta$-model with $r_{core}$=4.1~kpc and $\beta$=0.51, and we estimate the gas
mass from this model and the normalisation of the APEC model. The core of
NGC~835 is not well described by a single $\beta$-model (or even two), so
to reduce the bias arising from assuming a constant mean density in a
volume where density actually varies, we break it into two spectral
regions, a central cylinder of length $\sim$3.1~kpc and radius 0.97~kpc
aligned north-south, and the remaining 15\arcs\ (4.1~kpc) radius sphere.
For the diffuse emission in the disks of the two galaxies we approximate
the volume as a rectangular slab with projected area equal to that of the
polygon region used for spectral extraction, $\sim$13$\times$44~kpc and
depth 10~kpc. The third row of Table~\ref{tab:mgas} shows results summed
over all three regions in the NGC~833/835 complex. The gas density outside
the galaxy cores is 1.58$^{+0.85}_{-0.78}$$\times$10$^{-3}$\pcmcu.

\begin{deluxetable*}{lcccccccccc}
\tablewidth{0pt}
\tablecaption{\label{tab:mgas}Gas and stellar mass estimates}
\tablehead{
\colhead{Galaxy} & \colhead{log $L_K$} & \colhead{log $L_{3.6\mu m}$} & \colhead{$M_{\rm H\textsc{i}}$}& \colhead{$n_e$} & \colhead{$M_{\rm gas}^{\rm hot}$} & \colhead{$M_{*}^K$} & \colhead{$M_{*}^{3.6\mu m}$} & \colhead{$M_{\rm gas}^{\rm hot}$/$M_*^K$} & \colhead{$M_{\rm gas}$/$M_*^K$} \\
\colhead{} & \colhead{(\Lsol)} & \colhead{(\Lsol)} & (10$^9$\Msol) & \colhead{(10$^{-2}$\pcmcu)} & \colhead{(10$^8$\Msol)} & \colhead{(10$^{10}$\Msol)} & \colhead{(10$^{10}$\Msol)} & \colhead{($\times$10$^{-3}$)} & \colhead{($\times$10$^{-3}$)} \\
}
\startdata
NGC 833 & 11.06 & 10.99 & 0.79 & 0.80$^{+1.37}_{-0.76}$ & 0.61$^{+0.61}_{-0.18}$ & 7.46 & 4.83 & 0.62 & 11.21 \\[+.5mm]
NGC 835 & 11.27 & 11.28 & 1.17 & 2.15$\pm$1.29 & 0.64$\pm$0.36 & 12.10 & 9.62 & 1.03 & 10.70 \\[+.5mm]
NGC 833+NGC 835 & 11.48 & 11.45 & 1.96 &   -   & 3.57$^{+1.49}_{-1.27}$ & 19.56 & 14.45 & 1.83 & 13.19 \\[+.5mm]
NGC 838 & 10.95 & 11.17 & 3.02 & 2.27$^{+1.17}_{-1.31}$ & 2.57$^{+1.33}_{-1.48}$ & 5.79 & 7.40 & 4.43 & 56.60 \\[+.5mm]
NGC 839 & 10.92 & 11.03 & $>$4.47 & 1.86$^{+0.66}_{-0.69}$ & 0.41$\pm$0.15
& 5.41 & 5.36 & 0.76 & $>$83.38 \\[+.5mm]
NGC 848 & 10.67 & - & 0.77 & - & - & 3.04 & - & - & 25.33
\enddata
\tablecomments{We adopt $M_{*}$/$L_K$=0.65 and $L_{*}$/$L_{3.6\mu m}$=0.5
  \citep{Eskewetal12,McGaughSchombert14}. \Hi\ masses from
  \citet{VerdesMontenegroetal01}. No 3.6$\mu$m data are available for
  NGC~848, and we detect no hot gas in the galaxy.}
\end{deluxetable*}

In NGC~838 the diffuse emission is primarily located in the inner part of the galaxy or in the southern wind region. We approximate these two volumes as an oblate ellipsoid with radii 18.7\arcs\ and 10.1\arcs\ (5.1 and 2.76~kpc) and a conical section of a spherical shell with opening angle 43\degree\ and radii 25.5\arcs\ and 47\arcs\ (7.0 and 12.8~kpc). While the conical geometry of the wind in NGC~839 is obvious in H$\alpha$ \citep{Richetal10} in the X-ray it is less clear, and most of the wind and galaxy disk emission can be enclosed in a 15$\times$30\arcs\ box, suggesting a cylindrical geometry. 

Estimating the stellar mass of the major galaxies from their $K$-band and
3.6$\mu$m luminosities \citep[adopting stellar mass-to-light ratios
from][]{Eskewetal12,McGaughSchombert14} we find that, within the immediate boundaries
of the galaxies, the typical hot-gas-to-stellar mass ratio is
$\sim$10$^{-3}$, with the starburst superwind galaxy NGC~838 having the
highest ratio.  However NGC~835 contains the largest quantity of hot gas
(2.57$^{+1.33}_{-1.48}$$\times$10$^8$\Msol), perhaps unsurprising since it
has the highest stellar mass of the four. There is an order of magnitude
more cold \Hi\ than hot gas in the galaxies, giving total gas-to-stellar
mass ratios $\gtrsim$1\% for all five galaxies.

We can estimate approximate isobaric cooling times for the hot gas in the
galaxies, and find values of $\sim$1.4~Gyr and $\sim$1.8~Gyr for NGC~839
and NGC~838 respectively. This suggests that little of the gas in these
galaxy winds is likely to have cooled out of the X-ray regime. However, if
cooling via mixing with colder entrained gas dominates over radiative
cooling, the true cooling times will be shorter.

We estimate outflow rates for the two superwind galaxies based on the
geometries adopted above and an assumed outflow velocity. The only direct
measurement of outflow velocity comes from line splitting in the optical
spectra of NGC~839 which, accounting for the inclination of the galaxy and
opening angle of the conical H$\alpha$ wind structure, suggests a wind
velocity $\sim$250\kmps\ \citep{Richetal10}. This suggests an outflow of
$\sim$2.5\Msolpyr\ of hot gas in NGC~839. \citet{Vogtetal13} suggest, based
on the size of the wind-blown bubbles in NGC~838, that these structures are
5-50~Myr old with an outflow velocity of $\sim$130-1300\kmps. If we
assume a wind velocity similar to that in NGC~839, the bubbles would be
$\sim$27~Myr old. The outflow rate of hot gas is then $\sim$17\Msolpyr.

As star formation seems to have peaked in the two galaxies 400-500~Myr ago,
and declined since then, these rates are likely underestimates of the
average outflow over that period. However, if we assume outflows at these
rates over this timescale, we find that the two galaxies may have ejected
9.5$\times$10$^9$\Msol\ of hot gas. This value exceeds the \Hi\ mass of the
galaxies (and the hot gas mass in the surrounding region, see paper II),
suggesting either that the outflow rates are overestimated, or that gas
infall is replenishing the galaxies. The outflow velocity measured in
NGC~839 \citep[$\sim$250\kmps,][]{Richetal10} is greater than the escape
velocity of the galaxy \citep[$\sim$200\kmps, based on the stellar and
gaseous rotation velocities,][]{MendesdeOliveiraetal98} but only by
$\sim$20\%, suggesting that not all of the wind material will escape. The
two galaxies are also embedded in the \Hi\ filament, which is likely to
impede outflows and may be responsible for containing the bubbles seen in
NGC~838.  This may increase the fraction of outflowing gas which falls back
into the galaxy and is cooled and recycled. However, as we argue in paper
II, the morphology of the winds and of the surrounding diffuse emission
suggests that at least part of their outflows do escape and contribute to
the intra-group medium.

\subsection{Star Formation Rates}
\label{sec:starburst}

The star formation rates (SFRs) in the group member galaxies can be
estimated both from our X-ray and radio data, and from measurements in the
infra-red (IR) and ultra-violet (UV). The hard component of X-ray emission
in star forming galaxies arises primarily from high mass X-ray binaries,
formed from massive stars with short lifespans. It is therefore closely
linked to recent star formation, but not the current rate, since some time
is required for one member of each binary to evolve into a compact object.
\citet{Fragosetal13} suggest that the HMXB population is related to star
formation over the previous 2-100~Myr, but there is significant scatter in
the relationship. UV and IR measurements probe star formation via the most
massive, shortest-lived stars, and so may give a more accurate estimate of
the current star formation rate.

We use the $L_X$:SFR relation of \citet{Mineoetal12} to estimate star
formation rates from the 0.5-8~keV luminosity of the X-ray binary
population of each galaxy, as determined from the powerlaw component of the
galaxy emission from the best fitting models described in
section~\ref{sec:AGNdiff}, and including contributions from individual
point sources described in section~\ref{sec:PS}. We exclude the probable
background sources (numbers 11 and 16), and the nuclear sources (3, 12 and
14). Since individual ULXs may contribute a significant fraction of the
hard X-ray luminosity of an entire galaxy we calculate two star formation
rates for those galaxies containing ULX candidates, one including the ULX
contribution to luminosity, one with ULXs excluded. These estimates are
shown in Table~\ref{tab:SFR}.  In NGC~838 and NGC~839 the estimates are
complicated by the question of whether the galaxies host AGN. The evidence
suggests that source 3 in NGC~838 is probably an AGN, but if it were a ULX
it would increase the total SFR estimate for the galaxy to
16.68$^{+2.42}_{-2.30}$\Msolpyr. As we will argue in
Section~\ref{sec:disc839}, NGC~839 probably does not host an AGN, and we
base our SFR estimate on the AP+abs.PL model. If we instead assume an AGN
is present and use the luminosity of the unabsorbed powerlaw in the
AP+PL+abs.PL model to estimate SFR, we obtain a value of
8.74$\pm$0.74\Msolpyr.

To estimate star formation rates from the radio luminosity, we use the
$L_{1.4~GHz}$:SFR relation of \citet{Bell03}.  The GMRT 610~MHz data has
sufficient resolution for us to see that the AGN dominates the radio
emission in NGC~833 and for us to separate the central point source from
the extended star-formation emission in NGC~835. In NGC~838, the 610~MHz
image shows diffuse emission coincident with the wind bubbles north and
south of the galaxy, and we subtract this flux, which seems likely to be
associated with past rather than recent star formation.  Where necessary we
convert 610~MHz to 1.4~GHz fluxes using either the observed spectral index
(0.5-0.8 in NGC~838, NGC~839 and NGC~848) or a canonical value of 0.5 (for
NGC~835). The resulting SFR estimates are shown in Table~\ref{tab:SFR}. As
expected, NGC~838 and NGC~839 have the highest SFRs
(12.05$^{+2.50}_{-2.37}$ and 20.09$^{+2.06}_{-1.48}$\Msolpyr\
respectively), NGC~848 has a moderately enhanced rate of
8.76$^{+1.94}_{-1.78}$\Msolpyr, while NGC~833 and NGC~835 have rates of
$\sim$3\Msolpyr (all including ULXs).

\begin{deluxetable*}{lccccc}
\tablewidth{0pt}
\tablecaption{\label{tab:SFR}Star formation rate estimates}
\tablehead{
\colhead{Estimator} & \multicolumn{5}{c}{SFR (\Msolpyr)} \\
\colhead{} & \colhead{NGC~833} & \colhead{NGC~835} & \colhead{NGC~838} & \colhead{NGC~839}  & \colhead{NGC~848}\\
}
\startdata
X-ray (incl. ULXs) & 3.30$^{+0.43}_{-0.45}$ & 3.14$^{+1.26}_{-1.33}$ & 12.05$^{+2.50}_{-2.37}$ & 20.09$^{+2.06}_{-1.48}$  & 8.76$^{+1.94}_{-1.78}$ \\[+0.5mm]
X-ray (excl. ULXs) & 2.90$^{+0.43}_{-0.50}$ & 1.71$^{+1.28}_{-1.34}$ & 10.51$^{+2.51}_{-2.53}$ & - & - \\[+0.5mm]
Radio & - & 2.49$\pm$0.30 & 16.87$\pm$1.03 & 7.90$\pm$0.40 & 2.04$\pm$0.21 \\
SSP & 1-3 & - & 15-80 & - & - \\
24$\mu$m$^a$ & 0.07 & 1.74 & 7.24 & 9.55 & - \\
UV/IR$^b$ & 5.37$\pm$0.62 & 0.33$\pm$0.03 & 14.38$\pm$1.83 & 17.06$\pm$2.31 & -
\enddata
\tablecomments{$^a$: from Brassington et al. (2014, submitted).\\
$^b$: from \citet{Tzanavarisetal10}.}
\end{deluxetable*}

For comparison, we draw on two studies in the IR and UV:
\citet{Tzanavarisetal10}, who estimate SFRs using a combination of
\textit{Spitzer} 24$\mu$m and \textit{Swift} $uvw2$-band luminosities, and
Brassington et al. (2014, submitted), who use only \textit{Spitzer} data,
but use an updated $L_{24\mu m}$:SFR relation which is more accurate at the
high luminosities seen in starburst galaxies. Neither study includes
NGC~848. We also include estimates of the recent star formation rates from
the SSP models described in Section~\ref{sec:SSP}.

Although it is clear that the different estimators identify NGC~838 and
NGC~839 as the most actively star forming systems in HCG~16, there is a
significant degree of divergence among the actual SFR values. The SSP
estimate for NGC~833 is in general agreement with the other estimators,
though somewhat higher than the Brassington et al. value. The lower end of
the range of SFRs estimated from the SSP modelling of NGC~838 is in
agreement with the other estimates (though again higher than the
Brassington et al. result) but the upper limit greatly exceeds any of the
other estimates. Neither our BC03 or MILES model fits agree with the other
SFR estimates in detail; BC03 predicts current SFRs a factor $\sim$4.5
higher than the other estimators, while the MILES models suggest that star
formation ended $\sim$100~Myr ago. The best agreement is seen between the
Tzanavaris et al. estimates for NGC~838 and NGC~839, and our X-ray
estimates. However, the Tzanavaris rates are a factor $\sim$2 higher than
those found by Brassington et al., which should be more accurate. Removing
the contribution of ULXs brings our X-ray estimates into line with the
Brassington et al. estimates for NGC~835 and NGC~838. In NGC~839 our radio
estimate is closest to the Brassington et al.  estimate, while in NGC~838
it is closer to the Tzanavaris estimate.  Brassington et al. find that both
NGC~833 and NGC~835 have quite low star formation rates, 0.07 and
1.74\Msolpyr\ respectively.  The star forming ring in NGC~835 is probably
the main cause of the difference between the two galaxies. If the
Brassington et al.  estimates are correct, our X-ray estimates (and except
in the case of NGC~838, radio estimates) generally overpredict the current
SFR (particularly when ULXs are included) suggesting that all four original
compact group member galaxies were significantly more active in forming
stars over the past 10$^8$~yr than they are now.  This would agree with the
results of optical studies of NGC~838 and NGC~839
\citep{Vogtetal13,Richetal10}.

\subsection{Comparison with other star forming galaxies}
There is an extensive literature on star forming and starburst galaxies and
their winds, and comparison of NGC~838 and NGC~839 with population studies
offers the possibility of better understanding their physical state and
relationship to their environment. The sample of \citet{LiWang13} provides
a good basis for comparison. It consists of 53 nearby, highly inclined disk galaxies
selected to be star formation dominated with little or no AGN contribution,
all of which were observed with \chandra.

We follow Li \& Wang in estimating three further parameters: 1) the total
supernova mechanical energy injection rate, $\dot E_{SN}$. This includes type
Ia supernovae, with a rate estimated based on the stellar mass \citep{Mannuccietal05} as well as core collapse supernovae with a rate based on the SFR \citep{Heckmanetal90}, and assumes 10$^{51}$~erg per
supernova; 2) the X-ray radiation efficiency, $\eta$, defined as the
0.5-2~keV gas luminosity divided by $\dot E_{SN}$; 3) the surface rate of core
collapse (CC) supernovae, $F_{SN(CC)}$, defined as the number of supernovae
divided by the area of the galactic disk, determined from its \Dtf. We find
values of $\dot E_{SN}$=1.7$\times$10$^{42}$\ergps\ and $\eta$=2.4\% for
NGC~838, and $\dot E_{SN}$=2.4$\times$10$^{42}$\ergps\ and $\eta$=0.4\% for
NGC~839. The difference in X-ray radiation efficiency is interesting, and
is clearly caused by the larger mass of hot gas retained by NGC~838. Li \&
Wang show that NGC~838 has a high $\eta$, at the upper end of the range
seen in star-forming or starburst systems, and is among the most X-ray
luminous systems with comparable supernova energy injection rates. NGC~839
is apparently a more typical starburst.

The surface rate of CC supernovae provides information on the ability of
star formation to launch a wind out of the galactic disk.
\citet{Stricklandetal04b} showed that there is a critical value of
$F_{SN(CC)}$=25~SN~Myr$^{-1}$~kpc$^{-2}$ above which superbubble blowout
becomes possible. As expected, both our galaxies fall well above this
limit, with $F_{SN(CC)}\simeq$150~SN~Myr$^{-1}$~kpc$^{-2}$. However,
comparing these values with the $F_{SN(CC)}$:$\eta$ relation established by
Li \& Wang, we again find that NGC~838 is an outlier, with an efficiency
$\eta$ well above that expected for its surface rate of CC supernovae.
NGC~839 falls on the best-fitting relation. Both NGC~838 and NGC~839 have
relatively high gas temperatures for their X-ray luminosity and star
formation rate, at the outer edge of the scatter in these properties across
the galaxy population.

The morphology of the two galactic winds may provide an indication of the
cause of the difference in radiative efficiency between the two. In
NGC~838, the wind appears to be largely confined by the surrounding IGM,
forming two large (possibly leaking) bubbles. Conversely, the wind in
NGC~839 is conical and apparently unconfined. It seems plausible that
because of this confinement, NGC~838 has retained more gas and maintained a
higher X-ray luminosity. NGC~839 has a higher star formation rate, but a
lower hot gas content, suggesting that its wind has been able to escape the
galaxy, mixing with the surrounding IGM and cooling. This difference has
implications for the future development of the two galaxies, with NGC~838
potentially building up an enhanced metallicity and stellar fraction
compared to NGC~839. It is unclear why the NGC~838 wind is confined when
NGC~839 is not. Their specific star formation rates only differ by a factor
1.4-1.8 (depending on the band used to estimate stellar mass). Both
galaxies lie in the densest part of the \Hi\ filament, and of the hot IGM
(see paper II), and observationally this seems the most likely factor in
determining the difference in their properties.

\subsection{Does NGC~839 / HCG~16D host an active nucleus?}
\label{sec:disc839}
While early optical spectroscopic studies of NGC~839 found evidence of a
LINER or Seyfert nucleus \citep{DeCarvalhoCoziol99}, more recent
integral-field spectroscopy shows that the LINER-like line ratios are found
in an extended region and are more likely to arise from shock-excitation of
gas in the outflowing galactic superwind \citep{Richetal10}. However,
TRP01 classify the galaxy as hosting an absorbed AGN, based
on a spectral fit to the first light \xmms\ EPIC-MOS spectra which is
dominated by an absorbed powerlaw component at energies above 2~keV. Our
deep \chandra\ observations offer an opportunity to resolve this apparent
conflict.

We find that the \chandra\ spectra for the galaxy as a whole can be
described either by a thermal model with a single intrinsically absorbed
powerlaw, or by a combination of a thermal component, an unabsorbed
powerlaw representing the X-ray binary population, and an intrinsically
absorbed powerlaw representing an AGN. The two component model is a
marginally better fit than the three component model, but both are good
fits to the data (reduced $\chi^2$=0.991 or 1.039). The powerlaw indices in
both models are consistent with either AGN or X-ray binaries. Since we are
viewing HCG~839 almost edge-on ($i$=67\degree), emission from the central
regions of the galaxy is likely to be absorbed by dust and gas in the disk,
so the intrinsic absorption seen in the two component model could be
consistent with either an AGN or X-ray binaries.  As discussed in
Section~\ref{sec:starburst}, we can estimate the star formation rate in the
galaxy over the past $\sim$100~Myr based on the luminosity of the HMXB
population. The two component model implies
SFR=20.09$^{+2.06}_{-1.48}$\Msolpyr, while the three component model
suggests SFR=8.74$\pm$0.74\Msolpyr. The lower rate is more consistent with
the estimated rate of Brassington et al., based on the \textit{Spitzer}
24$\mu$m flux. The higher rate is closer to the estimated rate of
Tzanavaris et al., and since NGC~839 appears to have hosted more extensive
star formation in the past few 10$^8$~yr, with star formation rates
declining to their current level, a high X-ray star formation rate estimate
might be expected, owing to a larger population of high-mass X-ray
binaries.

Imaging the galaxy in the 2-7~keV band reveals a $\sim$5$\times$2.5\arcs\
ellipsoidal distribution aligned with the galaxy disk, with no clear
central point source. To model this hard band emission, we made a 2-7~keV
image from the combined 2013 observations, covering a $\sim$1\arcm\ region
centred on the galaxy. We convolved all models with an appropriate exposure
map and PSF. The image is well fitted using a constant background and an
elliptical $\beta$ model to represent the source emission. Best fitting
parameters were $r_{\rm core}$=2.34\arcs$^{+0.63}_{-0.40}$,
$\beta$=1.09$^{+0.25}_{-0.14}$ and ellipticity=0.47$\pm$0.04. If we add a
point source to represent an AGN and allow its position to vary freely, the
fit rejects the extra component, moving it out of the image. If the
position is fixed at the centre of the $\beta$ model, it provides
12.2$^{+22.4}_{-7.6}$ per cent of the 2-7~keV flux; even with a fixed
position, the point source contribution is consistent with zero at the
2$\sigma$ level.

In summary, the X-ray data alone do not conclusively rule out an AGN in
NGC~839, but suggest that the AGN contribution is probably weak if it is
present. Combined with the probability that the optical LINER emission
arises from shocks in the galactic superwind, this suggests that the
majority of spectrally hard X-ray emission in NGC~839 arises from its X-ray
binary population.

\subsection{Galaxy interactions as the trigger for nuclear activity and star formation}
The presence of the diffuse \Hi\ filament, combined with the \Hi\ deficiency estimates, demonstrates that tidal interactions have transported gas out of the four original group members. Based on the morphology, it seems plausible that a series of encounters produced a cloud of \Hi\ surrounding those four galaxies, which was then drawn out and elongated into a filament by the passage of NGC~848 through the group core. The motion of NGC~833/NGC~835 may also have helped extend the filament to the northwest. 

NGC~848 must have passed close to the other galaxies to have produced the
observed \Hi\ filament, and this raises the question of whether tidal
forces associated with its passage might be responsible for some of the
activity we observe. The estimated ages of the starbursts in NGC~838
\citep[500~Myr]{Vogtetal13} and NGC~839 \citep[400~Myr]{Richetal10} suggest
the possibility that these two starbursts were triggered in sequence, as
NGC~848 passed by from northwest to southeast. In such a scenario, the star
formation in NGC~848 could also have been triggered by these encounters.

It is less clear whether NGC~848 could have triggered star formation in
NGC~833 and NGC~835. The ongoing interaction between NGC~833 and NGC~835 is
probably sufficient to explain their nuclear activity. If we estimate a
velocity in the plane of the sky for NGC~848 based on the star formation
timescales, we find that we would expect it to have passed NGC~833
$\sim$600~Myr ago. A starburst of the magnitude seen in NGC~838 or NGC~839,
but 100~Myr older, should be clearly detected in the SDSS spectrum of
NGC~833, and ought to be obvious from optical colours in NGC~835.  However,
NGC~833 and NGC~835 are both $\sim$80\% \Hi\ deficient and are located in a
lower density region of the \Hi\ filament than NGC~838 and NGC~839. If both
galaxies lost most of their cold gas before the interaction, it is possible
that any triggered star formation was weak, short-lived, and largely
limited to NGC~835. Alternatively, NGC~848 may not have passed close enough
to them to trigger a starburst.

\section{Conclusions}
\label{sec:conc}

HCG~16 forms an excellent natural laboratory in which to study the effects
of star formation and nuclear activity driven by the tidal interactions
which are common in compact groups. Combining deep, high spatial-resolution
\chandra\ X-ray observations with VLA and GMRT radio data, we have examined
the five major galaxy members with the goal of determining the nature of
their present activity and its connection to the group environment. Our
results can be summarised as follows:

\begin{enumerate}
\item We identify eighteen point-like X-ray sources in and around the four original group member galaxies. Three of these are previously known sources associated with galactic nuclei, and one is probably a clump of gas in the starburst superwind of NGC~838. Two sources are located along the line of sight to background galaxies, consistent with our expectation of finding 2-3 background sources in the area of interest. The remaining twelve sources all appear to be associated with the group member galaxies, and five have luminosities $L_{0.5-8}\geq$10$^{39}$\ergps, making them candidate ultra-luminous X-ray sources (ULXs). One of the brightest point sources is found to be variable at $>$3$\sigma$ significance, is best described by a powerlaw spectral model with flat photon index ($\Gamma$=0.86$\pm$0.23), and is located $<$1\arcs\ from the optical centroid of NGC~838 (HCG~16C). We therefore suggest that this source may be a previously unidentified active nucleus.

\item We examine the X-ray spectra of NGC~833 and NGC~835 (HCG~16B and A)
  and confirm the previous finding that they host obscured AGN. SSP
  modelling of the SDSS spectrum of NGC~833 shows that little or no recent
  star formation has occurred in the galaxy. We use the excellent spatial
  resolution of \chandra\ to separate nuclear and galactic emission, and
  find that powerlaw emission previously thought to be scattered from the
  nuclei is more likely to arise from the X-ray binary populations of the
  two galaxies. The AGN components of both galaxies are found to be
  variable on timescales of months to years (between \chandra\
  observations), and we find evidence that this variability is best
  modelled as a change in the intrinsic luminosity of the AGN, and
  therefore probably of the accretion rate, rather than as a change in the
  optical depth of the absorber. We identify an Fe-K$\alpha$ emission line
  in NGC~835 with energy 6.41$\pm$0.05~keV and width 100$\pm$50~eV.

\item We find that NGC~838 and NGC~839 (HCG~16C and D) are probably both
  starburst-dominated systems, with minimal nuclear activity. Whereas
  previous X-ray studies have found evidence of an obscured AGN in NGC~839,
  in conflict with optical spectroscopy suggesting that LINER emission is
  produced by shocks in the galactic superwind, we find that the spectrally
  hard X-ray emission component is likely produced by the X-ray binary
  population of the galaxy and absorbed by gas and dust in the galactic
  disk. Both galaxies have mean temperatures $\sim$0.8~keV, and in
  NGC~838 we are able to constrain the abundance to
  0.16$^{+0.08}_{-0.04}$\Zsol.  However, this is likely biased low owing to
  mass loading of the wind with cold gas, and the wide range of
  temperatures in the hot phase of the wind. We examine the X-ray
  temperature structures of the winds and estimate the mass of hot gas
  involved and the rates of outflow. We also estimate the star formation
  rates of the galaxies from both X-ray and radio luminosities, and compare
  these to infra-red and ultra-violet estimates, finding that the X-ray
  rates probably better represent the SFRs in the recent past. The NGC~838
  wind appears to be confined by the surrounding \Hi\ filament and/or IGM,
  retaining more of its gas than has NGC~839. This has the potential to
  affect the development of the galaxy, with NGC~838 likely to have a
  higher metallicity and stellar fraction than NGC~839 in future.

\item We examine the X-ray and radio properties of NGC~848, the fifth largest galaxy in the group, using data from \chandra\ cycle~1, in which the galaxy falls $\sim$15\arcm\ off-axis. We find that the X-ray emission follows the optical bar and is best modelled by a simple powerlaw, suggesting that this starburst galaxy is dominated by emission from its X-ray binary population. The radio emission is more centrally concentrated, and may arise either from star formation or an AGN.

\end{enumerate}

\acknowledgements The authors thank L. Verdes-Montenegro for providing her
VLA \Hi\ map of the group, as well as C. Jones and the anonymous referee
for their helpful comments on the manuscript.  EO'S thanks R. Barnard for
useful discussions on data analysis. MT acknowledges the support of the Funda\c{c}\~{a}o de Amparo \`{a} Pesquisa do Estado de S\~{a}o Paulo (FAPESP),
process no. 2012/05142-5. Support for this work was provided by the
National Aeronautics and Space Administration (NASA) through Chandra Award
Number G03-14143X issued by the Chandra X-ray Observatory Center (CXC),
which is operated by the Smithsonian Astrophysical Observatory (SAO) for
and on behalf of NASA under contract NAS8-03060. SG acknowledges the
support of NASA through the Einstein Postdoctoral Fellowship PF0-110071
awarded by the CXC, and this research has made use of data obtained from
the Chandra Data Archive and software provided by the CXC in the
application packages CIAO, ChIPS, and Sherpa, as well as SAOImage DS9,
developed by SAO. We thank the staff of the GMRT for their help during
observations. GMRT is run by the National Centre for Radio Astrophysics of
the Tata Institute for Fundamental Research. We acknowledge the usage of
the HyperLeda database (http://leda.univ-lyon1.fr). Funding for SDSS-III
has been provided by the Alfred P. Sloan Foundation, the Participating
Institutions, the National Science Foundation, and the U.S.  Department of
Energy Office of Science.  The SDSS-III web site is http://www.sdss3.org/.

\textit{Facilities:} \facility{CXO} \facility{VLA} \facility{GMRT}

\bibliographystyle{apj}
\bibliography{../paper}

\end{document}